\documentclass[a4paper,UKenglish,cleveref, a4paper,autoref,thm-restate]{lipics-v2021}

\usepackage[dvipsnames]{xcolor}
\usepackage[boxruled,lined,linesnumbered]{algorithm2e}
\usepackage{amsthm,graphicx,graphics,amsmath,amssymb,enumitem}
\usepackage{comment}
\usepackage{url}
\usepackage[justification=centering]{caption}

\renewcommand\fbox{\fcolorbox{blue}{white}}
\hideLIPIcs  %uncomment to remove references to LIPIcs series (logo, DOI, ...), e.g. when preparing a pre-final version to be uploaded to arXiv or another public repository

\title{Engineering Semi-streaming DFS algorithms} %TODO Please add

\titlerunning{Engineering Semi-Streaming DFS algorithms} %TODO optional, please use if title is longer than one line

\author{Kancharla Nikhilesh Bhagavan}{Department of Computer Science and Engineering, Indian Institute of Technology Roorkee, India}{kancharla_nb@cs.iitr.ac.in}{}{}

\author{Macharla Sri Vardhan }{Department of Computer Science and Engineering, Indian Institute of Technology Roorkee, India}{macharla_sv@cs.iitr.ac.in}{}{}

\author{Madamanchi Ashok Chowdary}{Department of Computer Science and Engineering, Indian Institute of Technology Roorkee, India}{madamanchi_ac@cs.iitr.ac.in}{}{}

\author{Shahbaz Khan}{Department of Computer Science and Engineering, Indian Institute of Technology Roorkee, India}{shahbaz.khan@cs.iitr.ac.in}{https://orcid.org/0000-0001-9352-0088}{}

\authorrunning{
K. N. Bhagavan, M. S. Vardhan, M. A. Chowdary, S. Khan} %TODO mandatory. First: Use abbreviated first/middle names. Second (only in severe cases): Use first author plus 'et al.'

\Copyright{Kancharla Nikhilesh Bhagavan, Madamanchi Ashok Chowdary, Macharla Sri Vardhan, and Shahbaz Khan} %TODO mandatory, please use full first names. LIPIcs license is "CC-BY";  http://creativecommons.org/licenses/by/3.0/
 %TODO mandatory: Please choose ACM 2012 classifications from https://dl.acm.org/ccs/ccs_flat.cfm 

 %TODO mandatory; please add comma-separated list of keywords

\category{} %optional, e.g. invited paper

\relatedversion{} %optional, e.g. full version hosted on arXiv, HAL, or other respository/website
\supplement{Github Project: https://github.com/NikhileshBhagavan/DFS$\_$Semi$\_$Streaming}%optional, e.g. related research data, source code, ... hosted on a repository like zenodo, figshare, GitHub, ...
\funding{Startup Research Grant SRG/2022/000801 by DST SERB, Government of India. }%optional, to capture a funding statement, which applies to all authors. Please enter author specific funding statements as fifth argument of the \author macro.

\nolinenumbers %uncomment to disable line numbering

\EventEditors{John Q. Open and Joan R. Access}
\EventNoEds{2}
\EventLongTitle{42nd Conference on Very Important Topics (CVIT 2016)}
\EventShortTitle{CVIT 2016}
\EventAcronym{CVIT}
\EventYear{2016}
\EventDate{December 24--27, 2016}
\EventLocation{Little Whinging, United Kingdom}
\EventLogo{}
\SeriesVolume{42}
\ArticleNo{23}
\begin{document}

\maketitle

\begin{abstract}
Depth first search is a fundamental graph problem having a wide range of applications in both theory and practice. For a graph $G=(V,E)$ having $n$ vertices and $m$ edges, the DFS tree can be computed in $O(m+n)$ using $O(m)$ space where $m=O(n^2)$. In the streaming environment, most graph problems are studied in the semi-streaming model where several passes (preferably one) are allowed over the input, allowing $O(nk)$ local space for some $k=o(n)$. Trivially, using $O(m)$ space, DFS can be computed in one pass, and using $O(n)$ space, it can be computed in $O(n)$ passes.

Khan and Mehta~[STACS19] presented several algorithms allowing trade-offs between space and passes, where $O(nk)$ space results in $O(n/k)$ passes. They also empirically analyzed their algorithm to require only a few passes in practice for even $O(n)$ space. Chang et al. ~[STACS20] presented an alternate proof for the same and also presented $O(\sqrt{n})$ pass algorithm requiring $O(n~poly\log n)$ space with a finer trade off between space and passes. However, their algorithm uses complex black box algorithms, which are not suitable for practical implementations. 

We perform an exhaustive experimental analysis of the practical semi-streaming DFS algorithms. 
Our analysis ranges from real graphs to random graphs 
(uniform and power-law). We also present several heuristics to improve the state-of-the-art algorithms and study their impact. Our heuristics improve the state of the art by $\approx 40-90\%$, 
 achieving optimal one pass in almost $40-50\%$ cases (improved from zero). In random graphs as well, they improve from $30-90\%$, again requiring optimal one pass for even very small values of $k$. Overall, our heuristics improved the relatively complex state-of-the-art algorithm significantly, requiring merely two passes in the worst case for random graphs. Additionally, our heuristics made the relatively simpler algorithm practically usable even for very small space bounds, which was impractical earlier.
 \end{abstract}

\newpage

\section{Introduction}
\label{sec:Intro}

Depth first search (DFS) is a prominent graph algorithm having significant applications ranging from traversal~\cite{Tarjan72}, bi-connected components, strongly connected components, topological sorting~\cite{Tarjan76}, 
dominators~ \cite{Tarjan74}, etc. in directed graphs; and  connected components, cycle detection, 
edge and vertex connectivity~\cite{EvenT75}, 
bipartite matching~\cite{HopcroftK73}, planarity testing~\cite{HopcroftT74} etc. in undirected graphs. The DFS traversal computes a DFS tree according to the order of vertices visited during the traversal. This tree may not be unique since from a vertex any unvisited neighbour may be visited next. This non-uniqueness can be exploited to develop faster algorithms where DFS traversal does not follow a strict order to visit unvisited neighbours. On the other hand, lexicographically first DFS (LexDFS) creates a unique DFS tree by visiting the first unvisited neighbour in the adjacency list of a vertex. 
%Depth First Search explores all the edges of the graph so it also solves more basic problems
%such as computing connected components, detecting if the graph is bipartite, cycle detection
%(if the graphs is a DAG, in case of directed graphs), computing spanning tree, etc.
%DFS-tree is the spanning tree of the given graph formed by those edges through which
%vertices were visited for the first time.
In this paper, we address the problem of computing a general DFS tree in the semi-streaming environment. 

%In this paper, we address the problem of computing a DFS tree efficiently in the streaming environment. 
%In the semi-streaming environment, the input graph is accessed in form of a stream of graph edges,
%where an algorithm can perform multiple passes on this stream but is allowed to use only 
%$O(n)$ local space. Various important semi-streaming algorithms have been developed
%for some fundamental problems including ... Recently, Elkin~\cite{Elkin17} presented a semi-streaming
%algorithm to compute a shortest path tree in $n/k$ passes using $O(nk)$ space.

%In this paper we present an algorithm to compute a DFS tree in semi-streaming model. 
Space efficient computation have motivated studying various problems~\cite{FlajoletM85,HenzingerRR98,GuhaKS01,Indyk06} in the streaming model~\cite{AlonMS99,FeigenbaumKSV03,GuhaKS01}.  
The model requires the input to be accessed as a stream, over possibly multiple passes (preferably one) using only constant or \textit{logarathmic} local space.
In the case of graph problems, this limited space is found useful to compute only statistical properties (except computing triangles~\cite{Bar-YossefKS02}), making it impractical for other graph problems~\cite{FeigenbaumKMSZ05b,GuruswamiK16}. Hence, graph problems are typically studied under the more relaxed {\em semi-streaming model}~\cite{Muthukrishnan05,FeigenbaumKMSZ05}, which allows the storage size to $O(n~poly\log n)$ for a graph with $n$ vertices. Each pass sequentially reads the edges (possibly $O(n^2)$) where at no time the local space exceeds the above limit. In case the list of edges includes deletions, it is called \textit{turnstile} model else \textit{insertion-only}.
Various problems studied under this model are surveyed in ~\cite{ConnellC09,Zhang10,McGregor14}. While most results allow only a single pass over the input stream, the limitations of this restriction is also well studied~\cite{HenzingerRR98,BuchsbaumGW03,FeigenbaumKMSZ05,BorradaileMM14,GuruswamiO16}. Consequently, for various problems multi-pass algorithms~\cite{FeigenbaumKMSZ05,FeigenbaumKMSZ05b,McGregor05,AhnG13,Kapralov13,KaleT17} have been developed. Moreover, the allowed space of $O(n~poly\log n)$ have also been relaxed to $O(nk)$ (for $k=o(n)$) in~\cite{FeigenbaumKMSZ05b,Baswana08,Elkin11,Khan17,ChangFHT20}.

Computing a DFS tree is trivial using a single pass if $O(m)$ space is allowed, or using $O(n)$ passes if $O(n)$ space is allowed. The latter adds one vertex to the tree in each pass. Note that the former also works in \textit{turnstile} model. While several applications of DFS trees in undirected graphs have efficient solutions~\cite{WestbrookT92,FeigenbaumKMSZ05,FeigenbaumKMSZ05b,AusielloFL09,AusielloFL12,Farach-ColtonHL15,Kliemann16}, computing a DFS tree was considered an open problem~\cite{Farach-ColtonHL15,ConnellC09,Ruhl03} even for undirected graphs. Moreover, computing a DFS tree in ${O}(poly\log n)$ passes is considered hard~\cite{Farach-ColtonHL15}. Khan and Mehta~\cite{KhanM19} were the first to present non-trivial semi-streaming algorithms for computing DFS trees. Using $O(nk)$ space, they computed the DFS tree in $O(n/k)$ passes. It was further improved to $O(h/k)$ passes where $h$ is the height of the resultant tree where $h\leq n$. The above results were limited to \textit{insertion-only} model. Chang et al~\cite{ChangFHT20} presented an alternate proof for the $O(h/k)$ pass algorithm bound using sparse certificates for $k$ node connectivity. Further, they presented randomized algorithm for the \textit{turnstile model} requiring $\tilde{O}(n^3/p^4)$]\footnote{$\tilde{O}()$ hides the logarithmic factors, i.e., poly$\log n$ terms.} space for any $p\in[1,\sqrt{n}]$ passes. Also, for \textit{insertion-only} model, they presented an algorithm requiring $\tilde{O}(n^2/p^2)$ space. Note that for computing the DFS in $O(\sqrt{n})$ passes Khan and Mehta~\cite{KhanM19} required $O(n\sqrt{n})$ space while Chang et al.~\cite{ChangFHT20} requires $\tilde{O}(n)$ space. 

While the algorithm by Chang et al. ~\cite{ChangFHT20} uses complex black box algorithms, making it unsuitable for practical applications, Khan and Mehta~\cite{KhanM19} augmented their theoretical results with an experimental analysis for both random and real graphs, where their algorithms required merely a {\em few} passes even when the allowed space is just $O(n)$. 

%Elkin \cite{...} has recently shown how to compute single-source shortest path in $O(n/k)$ passes and
%$O(n.k)$ memory. In this paper we show a similar result for the computation of a DFS tree.
%We will show that it can be computed using $O(n/k)$ passes and $O(n.k)$ memory for any integer 
%$1\leq k\leq n$.
%%However, this bound is not very significant for most graphs whose DFS tree itself is know 
%We show a slight improvement in which we use $(h/k)$ passes and $O(n.k)$ memory where $h$ is the height
%of the resulting DFS tree.

%\subsection{Organisation}
%
%The paper is organized as follows. In section 2 we establish the terminology and notations
%used in the remainder of the paper. We also present a useful result in this secton. In section 3
%we begin by adapting the stored-data program for computing a DFS tree in the semi-streaming model 
%that requires $O(n)$ space and $n$ passes through edge-stream. We then improve it so that it requires 
%only $h$ passes where $h$ is the height of the computed DFS tree. In the subsequent sections we 
%present algorithms that require sublinear number of passes. Section 4 describes an algorithm that 
%builds the DFS tree $T$ by adding at least $k$ vertices in each pass, for any constant number $k$. 
%Hence it takes at most $n/k$ passes. This solution uses $O(n\cdot k)$ memory. Finally in section 5 
%we present the most advanced algorithm which grows $T$ by $k$ levels in each pass. Hence it takes 
%$h/k$ passes. This algorithm also requires $O(n\cdot k)$ memory.

\subsection{Related Work}
The semi-streaming algorithms for DFS trees have shown some intersection of ideas with dynamic graph algorithms, in particular considering the insertion of edges. Khan and Mehta~\cite{KhanM19} used a subroutine for updating a DFS tree on insertion of an edge from incremental DFS algorithm~\cite{BaswanaK17}. A semi-streaming algorithm for maintaining dynamic DFS of an undirected graph using $O(n)$ space was shown~\cite{Khan17} to require $O(\log^2 n)$ passes. Also, an experimental study on incremental DFS algorithms noted that the properties of DFS tree for random graphs allow computing the DFS tree in one pass using $\tilde{O}(n)$ space~\cite{BaswanaGK18}. 

On the other hand, if we exclude the memory required for storing the input and output, a trivial space-efficient algorithm using $O(n\log n)$ working space is known for computing the LexDFS tree. However, it was improved to require $O(n)$ working memory independently by Asona et al.~\cite{AsanoIKKOOSTU14} and Elmasry et al.~\cite{ElmasryHK15}. Given that LexDFS is a P-Complete problem, no polynomial time algorithms are possible, requiring $o(n)$ space for general graphs. However, Izumi and Otachi~\cite{IzumiO20} showed that it can be computed in $\tilde{O}(\sqrt{n})$ working space for special graphs as bounded tree-width graphs and planar graphs.

\subsection{Our Results}
In this paper, we explore the experimental evaluation of computing DFS in the semi-streaming model. We underscore the limitations of the existing state-of-the-art~\cite{KhanM19, ChangFHT20} and propose heuristics to improve them. We then compare the proposed algorithms with the existing algorithms on real and randomly generated graphs. For random graphs, we evaluate the algorithms on both uniform distribution and power law distribution. Recall that \cite{KhanM19} also performed the evaluation on real and random graphs (with uniform distribution only). 
Our results can be summarized as follows:
\begin{enumerate}
    \item \textbf{Corrections to state-of-the-art.} Previous practical algorithms~\cite{KhanM19} evaluates the use of $O(nk)$ space for constant values of $k$ resulting in inconsistencies. We thus define it explicitly by limiting the use of exactly $nk$ edges from the graph in the algorithm, which is more suitable for constant $k$. The corrected implementations and updated experimental results are thus presented.
    \item \textbf{Proposed heuristics.} An analysis of the state-of-the-art leads to several observations about the limitations of the existing algorithms. We thus proposed several heuristics to address these limitations. The heuristics essentially addressed the exploitation of pseudo root to save a pass, optimized use of the $nk$ edges, and further optimizations based on properties derived from related work. 
    \item \textbf{Evaluation on real and random graphs.} The algorithms with our proposed heuristics were evaluated against the state of the art in both real and random graphs. For real graphs, we improve the state-of-the-art by $\approx 40-90\%$  achieving optimal one pass in almost $40-50\%$ cases. In random graphs as well, the improvement ranges from $30-90\%$ requiring optimal one pass for a large segment of evaluated graphs. Overall, our heuristics improve the relatively complex state-of-the-art algorithm~\cite{KhanM19} by at least one pass, requiring only $2$ passes in the worst case for random graphs. However, the improvement in the relatively simpler algorithm~\cite{KhanM19} is exceptional for smaller values of $k$, making it a practical choice for all values of $k$ having comparable results as compared to the state-of-the-art. This was formerly untrue for very small $k$, where it was impractical.
\end{enumerate}

\section{Preliminaries}
\label{sec:prelim}
Given an undirected graph $G=(V,E)$ having $n$ vertices and $m$ edges, 
the DFS traversal of $G$ starting from any vertex $r\in V$ 
produces a spanning tree rooted at $r$ (or forest, in case $G$ is not connected),
called a DFS tree (or forest) in $O(m+n)$ time. For any rooted spanning tree of $G$, a {\it back edge} is a non-tree edge from a vertex to its ancestor, and {\it cross edge} is a non-tree edge that is not a back edge. 
A rooted spanning tree (or forest) is a DFS tree (or forest) of an undirected graph if and only if every non-tree edge is a back edge. The height of the computed DFS tree is denoted by $h$, which can be of any possible DFS tree and not necessarily the minimum or maximum height. 
%Let $G=(V,E)$ be any given undirected graph on $n=|V|$ vertices and $m=|E|$ edges. 
%\setlist[itemize]{noitemsep, topsep=3pt}
%\setlist[enumerate]{noitemsep, topsep=2pt}
Disconnected graphs and hence their DFS forests are handled by adding a dummy vertex $r$ to the graph
and connecting it to all vertices. The DFS tree rooted at $r$ in this updated graph has each child subtree of $r$ as a DFS tree of the DFS forest in the original graph. The following notations are used throughout the paper.
\begin{itemize}

\item $T:$ The DFS tree of $G$ is incrementally computed by our algorithm.
\item $par(w):$ Parent of $w$ in $T$. 
\item  $T(x):$ The subtree of $T$ rooted at vertex $x$. %[USED]
% \item  $path(x,y):$ Path from vertex $x$ to vertex $y$ in $T$.
% \item  $dist_T(x,y):$ The number of edges on the path from $x$ to $y$ in $T$.
% \item  $LCA(x,y):$ Lowest common ancestor of $x$ and $y$ in $T$.
\item $root(T'):$~ Root of a subtree $T'$ of $T$, i.e., $root\big(T(x)\big)=x$.
\item $level(v):$ Level of vertex $v$ in $T$, where $level(root(T))=0$ and
$level(v)=level(par(v))+1$.
\end{itemize}

%<<<<<<do we need this????>>>>>
%A subtree $T'$ is said to be {\em hanging} from a path $p$ if the $root(T')$ is a child of 
%some vertex on the path $p$ and does not belong to the path $p$, i.e., $T'\cap p=\emptyset$. 

The DFS tree $T$ will be built iteratively, starting from an empty tree. For any unvisited component $C$, the set of edges and vertices in the component will be denoted by $E_c$ and $V_c$. Further, each component $C$ maintains a spanning tree of the component that shall be referred to as $T_c$. $H_c$ of a component represents the union of spanning tree $T_c$ and the Edges of $E_c$ belonging to top k(some constant) levels of $T_c$. $E_c'$ term refers to the set of Edges belonging to $E_c$ that are stored during a pass to use them later.

\section{Previous work}
\label{sec:prevWork}
%We now briefly describe the previous results and their limitations.
\subsection{Khan and Mehta~\cite{KhanM19}}
The paper introduces several algorithms and heuristics, which are described as follows:

\begin{enumerate}
    \item \textbf{Simple \texttt{(Simp)}.}
In every pass, the algorithm adds an unvisited vertex to the DFS tree that has the lowest neighbour on the existing tree, requiring $\leq n$ passes. A beneficial heuristic adds a path instead of a single such vertex, resulting in fewer passes. 

\item \textbf{Improved \texttt{(Imprv)}.}
In every pass, the algorithm adds a vertex from each component of the unvisited graph, adding the next level of the DFS tree, requiring total $\leq h$ passes. % will constitute a separate subtree of the final DFS tree. Hence each such subtree can be computed independent of each other in parallel. The pass complexity of the algorithm is O(h) and the space complexity of O(m).

\item \textbf{k path \texttt{(kPath)}.}
In every pass, the algorithm uses $n'k$ edges for each component of $n'$ vertices and $m'$ edges to compute its DFS tree and add its longest path or entire DFS tree (if $m'\leq n'k$), adding at least $k$ vertices requiring $\leq \lceil n/k \rceil$ passes.
%In every pass of this algorithm, we store first Vc×K edges(where Vc  = number of vertices in component C, and K = space optimality factor) of every component and extract the longest path(or entire DFS tree if edges of the component are less than Vc×K edges) and add it to main DFS tree. The pass complexity of Kpath algo is O(n/k) and space complexity is O(n*k). The Pseudocode of this algorithm is attached in Figure (1) for reference.

\item \textbf{k level (\texttt{kLev}).} In every pass, the algorithm adds the next  $k$ levels of the DFS tree in each pass, using an incremental DFS algorithm as a subroutine~\cite{BaswanaK17}, requiring $\leq \lceil h/k \rceil$ passes. A beneficial heuristic additionally adds the vertices whose ancestors in the partial DFS tree are not modified during the pass, which has a big impact on the passes required.

%Thus, the algorithm requires exactly h/k passes. However, this algorithm also uses O(nk) space. Additional heuristics are introduced which adds unmodified vertices at the end of the pass to the DFS tree to reduce the pass complexity to less than h/k passes. We refer to this heuristic as marked/unmarked heuristic(H0) throughout this paper. The Pseudocode of this algorithm is attached in Figure (2) for reference. 

\end{enumerate}

In the experimental evaluation, both \texttt{kPath} and \texttt{kLev} performed much better than their theoretical bounds, requiring merely a \textit{few} passes even for $k=5,10$. While \texttt{kPath} required way more passes for smaller $k$, \texttt{kLev} had almost no impact of increasing $k$. Notably, \texttt{kPath} is also easier to explain and implement as compared to \texttt{kLev}. However, due to the ambiguous definition of space required, the implementation often ended up storing more than $nk$ edges of the graph, which required correction (see \Cref{apn:prevC} for details). Also, the following observations highlighted the scope of improvement in the existing implementations. \textit{Firstly,} the algorithms create a spanning tree using a complete pass despite having pseudo root $r$, which creates a trivial spanning tree. \textit{Secondly}, in several cases, despite having less than $nk$ edges to be processed, the algorithms required more than one pass due to implementation limitations. 
\textit{Thirdly,} Baswana et al.~\cite{BaswanaGK18} previously highlighted a heuristic that allowed computing a DFS tree of random graphs in a single pass using $O(n\log n)$ space, which highlighted those edges to be ignored which cannot affect the residual uncomputed DFS tree. We thus propose several new heuristics to improve the state-of-the-art algorithms \texttt{kLev} and \texttt{kPath}.

\subsection{Chang et al.~\cite{ChangFHT20}} 
Chang et al.~\cite{ChangFHT20} improved the theoretical results by Khan and Mehta~\cite{KhanM19} allowing $O(\sqrt{n})$ passes using $\tilde{O}(n)$ space. They also give smooth tradeoffs between space required and number of passes as follows. For computing the solution in $p\in [1,\sqrt{n}]$ passes they presented a randomized algorithm requiring $\tilde{O}(n^3/p^4)$ passes allowing both the more flexible turnstile model having possible deletions in the stream. They also give a deterministic algorithm requiring $\tilde{O}(n^2/p^2)$ passes for insertion only model. Their approach relies on a semi-streaming implementation of parallel algorithm to compute DFS by Aggarwal and Anderson~\cite{AggarwalA88} and sparse certificates for $k$ node connectivity~\cite{EppsteinGIN97,GuhaMT15}. 

However, despite having better theoretical bounds, they use several complex black box algorithms and data structures, which limits their application in practice.

%The harder part in this approach is computing  maximal node disjoint paths between a set of vertices. 

%The state- current best theoretical algorithm is by chang~\cite{ChangFHT20} which is obtained by streaming implementation of the Algorithm of Aggarwal and Anderson. The bottleneck of this algorithm is a task called MaximalPaths which is a variation of maximal node disjoint paths problem between a set of source nodes and a set of sink nodes. In Aggarwal and Anderson algorithm this MaximalPaths problem should be implemented parallelly which in further should be implemented in streaming environment which makes it practically harder to implement. Also expermental results of Algorithms by Khan have shown exceptionally better results. So, in this paper we have focused on improving the results of Algorithms by Khan by adding some heuristics.

\section{Proposed Heuristics}

Based on the above observations of Khan and Mehta~\cite{KhanM19}, we propose the following heuristics (pseudocodes for the same are presented in \Cref{apn:pseudocodes}).

\subsection*{Heuristic \texttt{H1}: Exploiting Artificial root}

In both \texttt{kPath} and \texttt{kLev} algorithms, each component is processed independently, requiring the computation of its spanning tree. Thus, in the previous implementation the first pass is used for computing this spanning tree being different than the remaining passes. However, using the artificial root $r$ (recall \Cref{sec:prelim}) to make the DFS forest a tree, we get a trivial spanning tree where each node is a child of $r$. Thus, this modification simplifies the implementation (no distinct first pass) and avoids wasting a pass. Notably, the minimum passes required for \texttt{kPath} and \texttt{kLev} were previously $2$, which are now reduced to  $1$, which is optimal. 

%We can observe that in the KPath and KLev algorithms of Khan and Mehta, the component is processed only after the computation of its spanning tree. In the implementation of these algorithms, the first pass is just getting used for adding an artificial vertex('0') to DFS tree as root, and thereafter finding remaining components and spanning trees of them which will be processed in the next passes. Instead of wasting the whole pass for just computation of components and spanning trees, we can always create a spanning tree at the beginning of the algorithm, just by adding edges from artificial vertex('0') to each vertex with artificial vertex as root. This optimises the first pass that was not been used for processing any components.

\subsection*{Heuristic \texttt{H2:} Optimizing $nk$ edge utilization level 1}

The previous implementations of \texttt{kPath} and \texttt{kLev} wastefully employed the 
$nk$ edges allowed to be stored, limiting their practical performance. These can be improved as follows.

\begin{itemize}
    \item \texttt{kPath:} Each pass stores a spanning tree for each component having $n'$ vertices along with its first $n'(k - 1)$ edges giving $n'k$ edges, and overall $nk$ edges. However, some of these edges might already have been in the spanning tree. Removing such duplicates will help us collect all $n'k$ unique edges, guaranteeing each component has the longest path of length $\geq k$. 

\item \texttt{kLev:}
Each pass stores all the non-tree edges (back edges) to top $k$ unvisited levels, resulting in the correct computation of top $k$ levels and maximum $nk$ edges. However, many of these edges may not exist, resulting in storing much lower number of edges. This can be improved by storing edges to top $i (\geq k)$ levels as long as their number is $\leq nk$, allowing adding top $i$ levels instead. This can be performed by storing all vertices by level and removing back edges to lower vertices while maintaining $i$. 
\end{itemize}

\subsection*{Heuristic \texttt{H3}: Optimizing $nk$ edge utilization level 2}

\begin{itemize}
\item \texttt{kPath:} If we carefully analyze after the first pass, we are utilizing storing $n^*k$ edges instead of $nk$ edges, where $n^*$ are the number of unvisited vertices in the computed DFS tree. We thus define $k_{opt}$ as $nk/n^*\geq k$. We can thus process each component using $n'k_{opt}$ edges, potentially exploring more edges leading to a longer path.

\item \texttt{kLev:} Baswana et al.~\cite{BaswanaGK18} highlighted a broomstick property where, for random graphs, the length of the top path in a DFS tree without branching grows rapidly. The back edges to this path do not affect the DFS tree computation as this DFS tree becomes fixed, allowing us to ignore these edges utilizing $nk$ edge bound more efficiently. Marking the lowest vertex of the broomstick and ignoring the back edges for the vertices above marked vertex will save the space which can be used by other vertices increasing the number of levels added in each pass.
\end{itemize}

%This heuristic is only for Klev algorithm proposed in khan and Mehta paper. Top path heuristic leverages the broom stick property, that most of the random graphs and few real graphs exhibit to effectively use the nk space available for storing backedges. In graphs with this property, a Depth-First Search (DFS) traversal tends to encounter long "broom stick" branches before reaching the rest of the graph. These branches extend from a central node, resembling the handle of a broom, with numerous shorter branches or leaves attached to them. Upon observation, it's clear that storing the backedges of all the vertices constituting the long broomstick isn't beneficial, as these vertices won't fall any further in the future. So, marking the lowest vertex belonging to this broom stick would be enough. We do this by initializing the artificial root (to which all the vertices are attached initially) as the marked vertex. As the algorithm progresses and the DFS tree is computed, we drop this marked vertex down to the vertices below whenever the count of the number of children for the marked vertex becomes one and delete the backedges stored for all vertices above the marked one.

\noindent
\textbf{Remark:} We also explored other heuristics, such as randomizing the existing adjacency list before computing the longest path in \texttt{kPath} as random graphs have larger longest paths~\cite{BaswanaGK18}. However, there was no visible improvement; hence, we have not included it in our analysis. 

\section{Experimental Setup}
We now evaluate the state-of-the-art practical algorithms using existing and newly proposed heuristics on real and random (uniform and power law) graphs. For real graphs, we use the same datasets previously used~\cite{KhanM19} except those that are no longer publicly available. For random (uniform and power law) graphs, we use Erd\H{o}s R\'{e}nyi $G(n, m)$ model~\cite{Bollobas84,ErdosR60} and power law~\cite{em/999188420} distribution property respectively varying one of the parameters vertices $n$, edges $m$, or allowed stored edges parameter $k$ at a time, similar to~\cite{KhanM19}. 

%For random graphs, they limited the evaluation to three variations on uniform random graphs. However, uniform random graphs do not simulate the real world data accurately, which is why it is more suitable to additionally evaluate the data on power law graphs. 

\subsection{Algorithms and Implementation details}\label{section:base-algos}
We evaluate the following algorithms:
\begin{enumerate}
    \item \texttt{kPathO.} The previous algorithm~\cite{KhanM19} requiring $O(n/k)$ passes for $O(nk)$ space. 
    \item \texttt{kPath1.} The proposed algorithm using \texttt{H1} heuristic to improve \texttt{kPathO}.
    \item \texttt{kPath2.} The proposed algorithm using \texttt{H1} and \texttt{H2} heuristics to improve \texttt{kPathO}.
    \item \texttt{kPathN.} The proposed algorithm using all additional heuristics to improve \texttt{kPathO}. \item \texttt{kLevO.} The previous algorithm~\cite{KhanM19} requiring $O(h/k)$ passes for $O(nk)$ space.
    \item \texttt{kLev1.} The proposed algorithm using \texttt{H1} heuristic to improve \texttt{kLevO}.
    \item \texttt{kLev2.} The proposed algorithm using \texttt{H1} and \texttt{H2} heuristics to improve \texttt{kLevO}.
    \item \texttt{kLevN.}  The proposed algorithm using all additional heuristics to improve \texttt{kLevO}.
\end{enumerate}

Khan and Mehta~\cite{KhanM19} analyzed the performance of the algorithms for different values of $k=1,2,5,10$. However, for such constant values of $k$, allowing $O(nk)$ space becomes ambiguous because of the additional data structures and variation of $k$ by constant factors. Hence, it is more suitable to define the space limitation in terms of the number of edges of the graph stored at any time to be $\leq nk$. This warrants correcting the previous algorithms for a fairer comparison under the new restrictions, as described in \Cref{apn:prevC}.

\subsection{Evaluation Metrics}
The primary metric to evaluate the semi-streaming algorithms is the number of passes required and the space utilized. Since the use of $O(nk)$ space is enforced by limitation to store $\leq nk$ edges from the graph, the only significant metric required is the number of passes. The performance in terms of time taken by the algorithms (and heuristics) is not emphasized as the focus was not on the most efficient implementation of the heuristics, but rather on their impact to reduce the number of passes.

\subsection{Environment Details}

%\noindent\textbf{GitHub Repository.} \href{https://github.com/Bashar-Ahmed/Safe-Flow-Decomposition}{https://github.com/Bashar-Ahmed/Safe-Flow-Decomposition}

All the evaluated algorithms are implemented in C++ and compiled using GNU C++ compiler version 11.4.0 using -O3 optimization flag for compilation.  
%All C++ implementations use optimization level 3 of GNU C++ (compiled with -O3 flag). 
% The performance of the algorithms was evaluated on the PARAM Ganga\footnote{https://hpc.iitr.ac.in} supercomputer system at IIT Roorkee, using a single core of Intel Xeon Platinum 8268 CPU 2.90GHZ with 16GB memory 
The performance of the algorithms was evaluated using the WSL(Windows subsystem for Linux) on laptops
%, using a four cores of 
having Intel(R) Core(TM) i5-10210U CPU $@$ 1.60GHz with 8 GB RAM.
% \sbzk{?? 192?}. 
%The source code of our project is available on GitHub\footnote{https://github.com/NikhileshBhagavan/DFS$\_$Semi$\_$Streaming} under GNU General Public License v3 license.  %flag in an 8 GB RAM system. 

% \textbf{UPDATE THIS}
%To remove noise and other disturbances from affecting the results, we run each algorithm 25 times, measuring for each run, the required metrics. Further, we clip the outliers and take the average value as our required metric.

\subsection{Datasets}\label{sec:datasets}

In our experiments, we considered the following types of datasets.
\begin{enumerate}
\item \textbf{Real graphs:} We use the following publicly available undirected graphs sourced from real-world instances in the KONECT database~\cite{Konect13}. These can be classified in following types, namely, online social networks (HM~\cite{PHamster16}, BrightK~\cite{ChoML11}, LMocha~\cite{ZafaraniL09}, FlickrE~\cite{McAuleyJ12}, Gowalla~\cite{ChoML11}), human networks (AJazz~\cite{PabloL03}, ArxAP~\cite{LeskovecKF07}, Dblp~\cite{LeskovecY12}), recommendation networks (Douban~\cite{ZafaraniL09}, Amazon~\cite{LeskovecY12}), infrastructure (CU~\cite{Knuth08}),  autonomous systems (AsCaida~\cite{LeskovecKF07}), lexical words (Wordnet~\cite{Fellbaum98}). For  relative sizes see \Cref{tab:reduce-pass,tab:KPathold_KPathnew_RedPer,tab:KLevold_KLevnew_RedPer} in Appendix.\\

\item \textbf{Random Graphs:} We have generated the random graphs using the uniform distributions. In these graphs, the update sequence is generated based on Erd\H{o}s R\'{e}nyi $G(n, m)$ model~\cite{Bollobas84,ErdosR60} by choosing the first $m$ edges of a random permutation of all possible edges. \\

\item \textbf{Power-Law Graphs \cite{em/999188420}:} We have generated the power-law graphs using the property of power-law distribution. These graphs are generated by iteratively adding edges between nodes selected based on probabilities proportional to the exponent of their degrees. The power law exponent used for generating the graphs is $3$.

\end{enumerate}

\section{Results}
\subsection{Real Graphs}
% \Cref{fig:realImprov} shows the improvement from \texttt{kPathO} to \texttt{kPathN} and \texttt{kLevO} to \texttt{kLevN} for various datasets with respect to changing space parameter $k$. Detailed results of improvements for each datasets are described in \Cref{apn:detailReal}. Overall, \texttt{kLevN} performs better than all the algorithms where difference with \texttt{kPathN} is more for smaller values of $k$.

\begin{figure}[h]
  \begin{subfigure}{.5\textwidth}
  \centering
    \includegraphics[trim={0 1cm 0 0},clip,width=\linewidth]{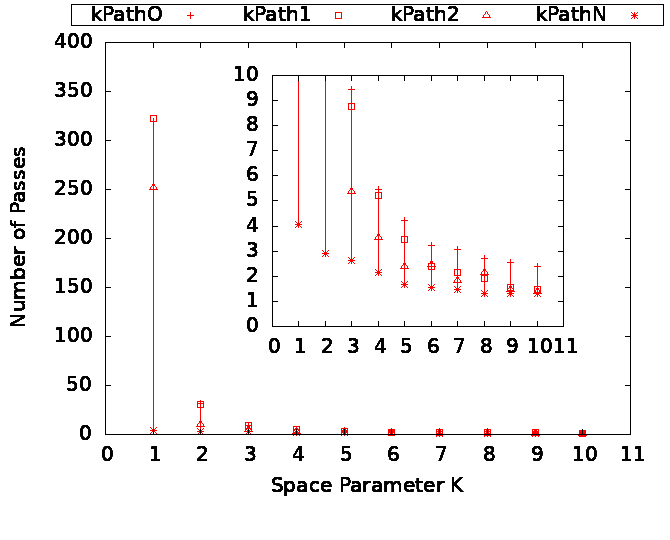}
  \end{subfigure}%
  \begin{subfigure}{.5\textwidth}
  \centering
    \includegraphics[trim={0 1cm 0 0},clip,width=\linewidth]{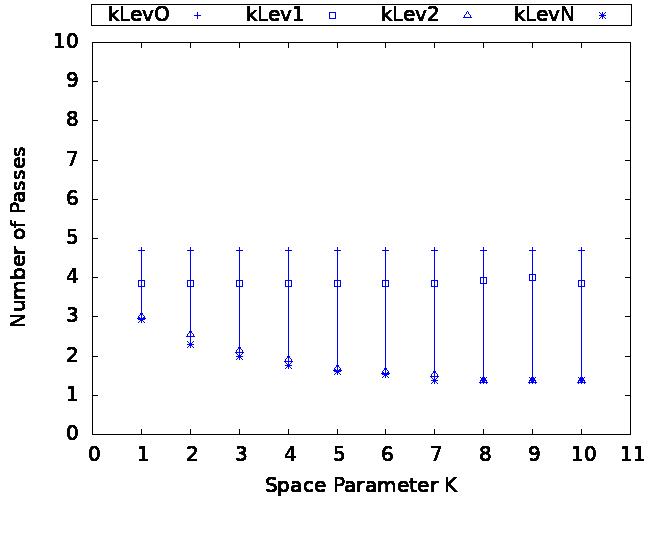}
  \end{subfigure}
  \caption{Average improvement in required passes for  \texttt{kPath} and \texttt{kLev} for real datasets.}
  \label{fig:realImprov}
\end{figure}

\Cref{fig:realImprov} summarizes the results showing the average improvement over all the datasets from \texttt{kLevO} to \texttt{kLev1}, \texttt{kLev2} and \texttt{kLevN}, and of \texttt{kPathO} to \texttt{kPath1}, \texttt{kPath2}, and \texttt{kPathN} with respect to changing space parameter k. The details of the datasets and the detailed results of improvements for each dataset are described in \Cref{apn:detailReal}. Overall, the heuristics improve state of the art significantly, where \texttt{kLevN} always performs better than \texttt{kPathN}.

The key thing to note in \texttt{kPath} algorithms is the improvements obtained from \texttt{kPath1} to \texttt{kPathN} are enormous over the improvements obtained from \texttt{kPathO} to \texttt{kPath1}, which is more ($65-90\%$) for smaller values of $k\geq 3$ than higher values $k\geq 4$ ($45-55\%$). The average improvement of \texttt{kPathN} over \texttt{kPathO} is massive($\approx90\%$) for smaller values of $k=1$. This is because the wastage of space happening by utilizing $n^*k$ instead of $nk$ (recall \texttt{H3}) in the \texttt{kPathO} algorithm is more for smaller values of $k$. Therefore, correcting this flaw in \texttt{kPathN} has shown massive improvements for smaller values of $k$. For higher values of $k$, since existing passes are less, even \texttt{H1} contributes significantly ($40-60\%$). Notably, several results ($\approx40\%$) now show the requirement of optimal \textit{one} pass showing the impact of \texttt{H1}, given in such cases usually, we have $m < nk$. From the graph for \texttt{kPath} we can also observe that \texttt{H3} contribution is much more than the contribution made by \texttt{H1} and \texttt{H2} for smaller values of $k$, and with the increase in $k$ value, contribution by \texttt{H3} keeps decreasing while the contribution by \texttt{H1} and \texttt{H2} keeps increasing. From the \Cref{tab:KPathold_KPathnew_RedPer} of \Cref{apn:detailReal}, we can observe a rough trend where the improvement from \texttt{kPathO} to \texttt{kPathN} is increasing with the increase in value of $m$. In a few cases, we can see \texttt{kPath1} taking more passes than \texttt{kPathO}, which can be attributed to different residual graphs which can lead to such anomalies.

The average improvement of \texttt{kLevN} over \texttt{kLevO} is less ($30-50\%$) for smaller values of $k$($\leq 3$) than higher values $k\geq 4$ ($60-70\%$). Notably, the performance of \texttt{kLevO} was relatively independent of $k$, but with the addition of \texttt{H2} heuristic, \texttt{kLevN} better exploits the allowed $nk$ edges. This reflects in more improvement for higher $k$ values. Also, for many cases ($\approx 50\%$), we now achieve the result in one optimal pass (see \Cref{apn:detailReal}), which was \textit{two} earlier. An additional thing to note is that the drop in the number of passes by the addition of the \texttt{H1} heuristic is nearly \textit{one} in most cases, but a huge drop occurs with the addition of the \texttt{H2} heuristic. As the \texttt{H2} heuristic successfully exploits the efficient use of $nk$ space, the effect of adding the \texttt{H3} heuristic is not too much in the case of real graphs.

%for smaller k, so although the improvement in the number of passes is high the percentage of improvement is relatively less. We can observe that for the datasets ArxAP and FlickrE, kPathO is performing better than kPathN which is an anomaly. For KPath, in figure [\ref{fig:kpath-new-real}] we can observe that after a certain k value for kPathN, the number of passes required to compute the DFS tree has become 1. This is because if the total number of edges is less than n(k-1) then we can directly compute the DFS tree as there is no space constraint. 

\begin{observation}
On evaluation over real datasets, we have:
\begin{enumerate}
    \item \texttt{kLevN}  
    outperforms all algorithms in almost all cases. \item \texttt{kPathN} and \texttt{kLevN}  improves respectively more over \texttt{kPathO} and \texttt{kLevO} for lower $k$ ($\approx 80\%,40\%$), than higher $k$ ($\approx 50\%,65\%$).
    \item For \texttt{kPath}, \texttt{H1} improves by a pass, while impact of \texttt{H2} is relatively same $\approx20\%$, and that of \texttt{H3} is maximum (up to $98\%$) though sharply decreases with $k$.
    \item For \texttt{kLev}, \texttt{H1} improves by a pass, while impact of \texttt{H2} increases with $k$ up to $\approx60\%$, and that of \texttt{H3} is negligible.
    \item Optimal \textit{one} pass is achieved by \texttt{kPathN} ($\approx 40\%$) and \texttt{kLevN} ($\approx 50\%$). 
\end{enumerate}
\end{observation}

\subsection{Random Uniform Graphs}
\begin{figure}[!h]
  \begin{subfigure}{.5\textwidth}
  \centering    \includegraphics[trim={0 1cm 0 0},clip,width=\linewidth]{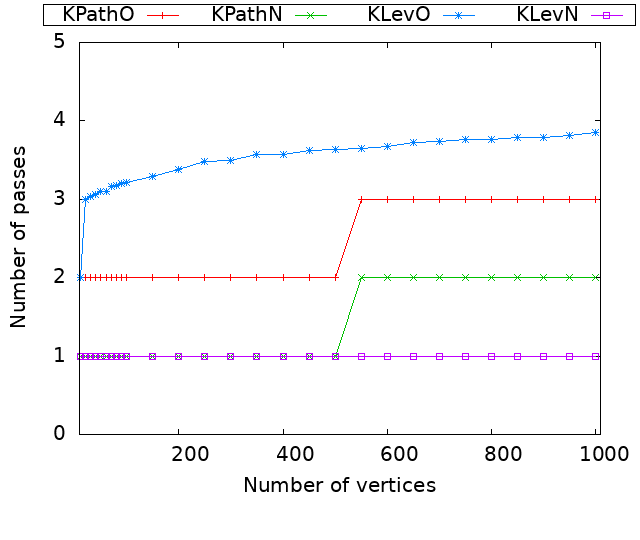}
    \caption{Density $m=O(n\log n)$}
  \end{subfigure}%
  \begin{subfigure}{.5\textwidth}
  \centering
    \includegraphics[trim={0 1cm 0 0},clip,width=\linewidth]{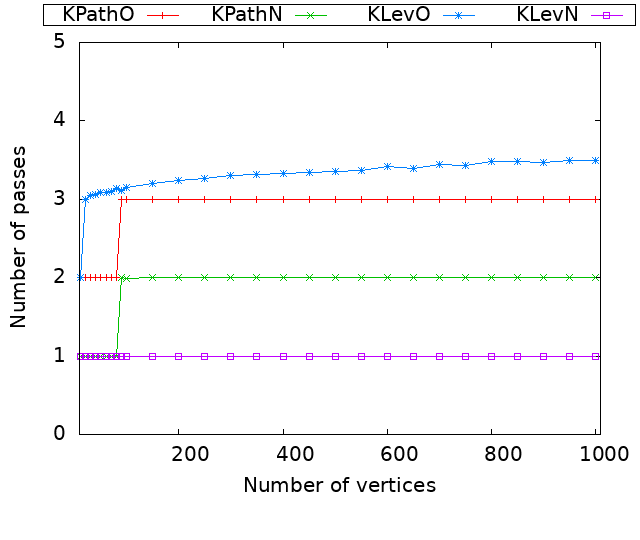}
    \caption{Density $m=O(n\sqrt{n})$}
  \end{subfigure}
  \caption{
Performance of the faster algorithms as the number of vertices is varied with (a) $m=O(n\log n)$, and (b) $m=O(n\sqrt{n})$ densities, using $10n$ edges ($k=10$).}
  \label{fig:randVarN}
\end{figure}

\begin{figure}[!h]
  \begin{subfigure}{.5\textwidth}
  \centering
    \includegraphics[trim={0 1cm 0 0},clip,width=\linewidth]{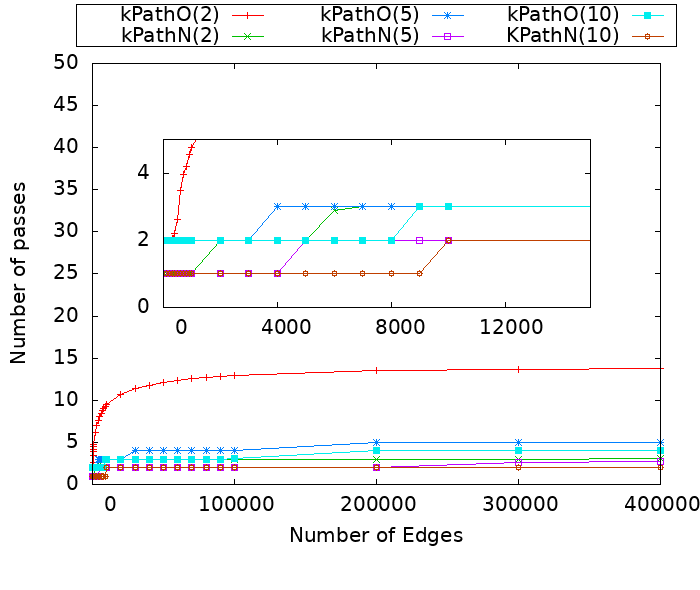}
    \caption{Improvement in \texttt{kPath}}
  \end{subfigure}%
  \hfill
  \begin{subfigure}{.5\textwidth}
  \centering
    \includegraphics[trim={0 1cm 0 0},clip,width=\linewidth]{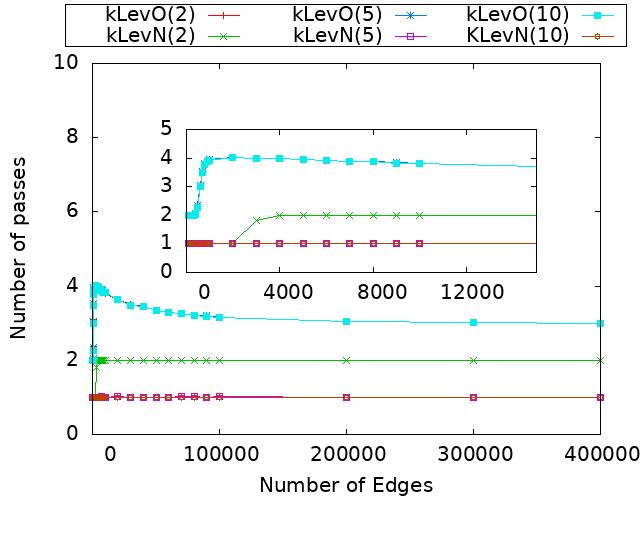}
    \caption{Improvement in \texttt{kLev}}
  \end{subfigure}
  \caption{Performance of the algorithms as the number of edges are varied up to $O(n^2)$ for $n=1000$ vertices and different values of allowed edges ($k=2,5,10$) for (a) \texttt{kPath}, and (b) \texttt{kLev}.} % VarM for n = 1000 and m$\_$max = nC2 for K = 5,10 }
    \label{fig:randVarM}
\end{figure}

\begin{figure}[!h]
  \begin{subfigure}{.5\textwidth}
  \centering
    \includegraphics[trim={0 1cm 0 0},clip,width=\linewidth]{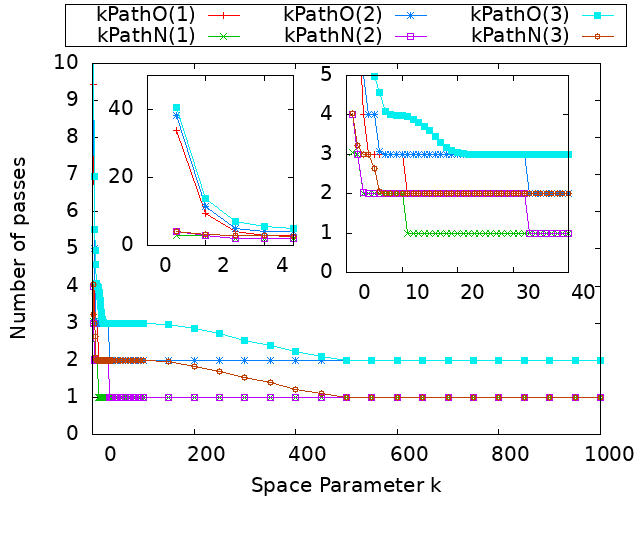}
    \caption{Improvement in \texttt{kPath}}
  \end{subfigure}%
  \begin{subfigure}{.5\textwidth}
  \centering
    \includegraphics[trim={0 1cm 0 0},clip,width=\linewidth]{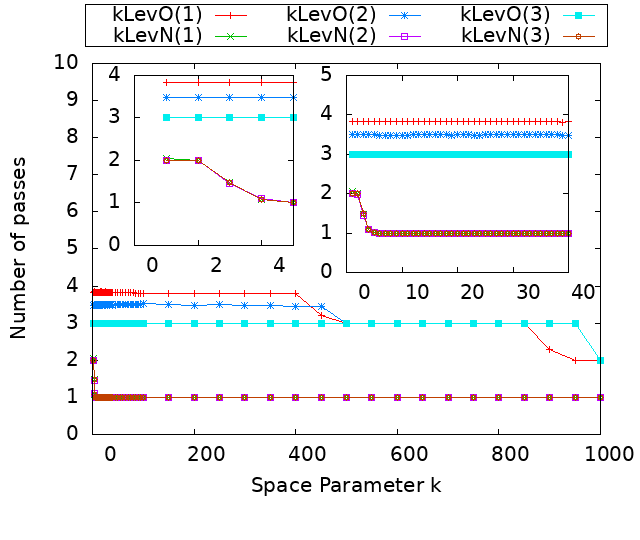}
    \caption{Improvement in \texttt{kLev}}
  \end{subfigure}
 \caption{Performance of algorithms as the number of stored edges are varied up to $k=n$ for $n=1000$ and different values of $m=O(n\log n), O(n\sqrt{n}),O(n^2)$ for (a) \texttt{kPath}, and (b) \texttt{kLev}.} %VarK for n = 1000 and (1) = nlogn (2) = n$\sqrt{n}$ (3) = nC2}
\label{fig:randVarK}
\end{figure}

In random graphs, we perform three kinds of experiments evaluating the performance of the algorithms by varying the three parameters: number of vertices $n$ (\Cref{fig:randVarN}), the edge density $m$ (\Cref{fig:randVarM}) and allowed stored edges parameter $k$ (\Cref{fig:randVarK}). However, we do not explore the impact of each heuristic separately but rather focus on the overall impact. 

Overall, \texttt{kLevN} always performs the best among the algorithms. The impact of \texttt{H1} is clearly visible for both \texttt{kLev} and \texttt{kPath}, reducing the number of passes by at least one. 
Also, both the new algorithms require only one pass (down from 2) when the edges $m\leq n(k-1)$. This is because our algorithm adds $n$ extra edges to make the initial spanning tree from the artificial root before the first pass. Thus, if the total effective edges are $\leq nk$ edges, the DFS tree is computed in the first pass itself.

In \Cref{fig:randVarN}, we evaluate the number of passes required for a graph with varying $n$ having the number of edges $m$ as $O(n\log n)$ (left) and $O(n\sqrt{n})$ (right). We observe that the algorithm \texttt{kLevN} again outperforms the other algorithms significantly and \texttt{kLevO} by $66-75\%$. \texttt{kPathN} reduces one pass over \texttt{kPathO} effectively improving by $33-50\%$. The exceptional performance of \texttt{kLevN} requiring one pass for all cases can be attributed to the broomstick structure (exploited by \texttt{H3}), thus having expected performance~\cite{BaswanaGK18} (since $k=10\geq \log n$). Note that Baswana et al.~\cite{BaswanaGK18} proved that the DFS tree of a random graph can be computed in a single pass using $O(n\log n)$ space due to the broomstick structure.

%Using the H3 heuristic \texttt{kLevN} adds the longest path, and the bristles will be added directly using the main algorithm.

In \Cref{fig:randVarM}, we evaluate the number of passes required for fixed $n=1000$ and varying $m$ up to $O(n^2)$ for different values of $k=2,5,10$. %Note that in both \Cref{fig:randVarN,fig:randVarM} we can observe that the \texttt{kPath} algorithm takes only one pass (\texttt{kPathO} taking 2 passes) in all the cases where the edges in the graph are $\leq n(k-1)$. This is because our algorithm adds $n$ extra edges for the sake of artificial root before the first pass, when the total edges are less than $nk$ edges then the DFS tree of the total graph will be computed and added in the first pass itself. 
In addition to saving a pass due to \texttt{H1}, the difference in step from  2 to 3 for \texttt{kPathO} as compared to 1 to 2 for \texttt{kPathN} can be attributed to the duplicate edges removed from \texttt{kPathN} (\texttt{H1}). This effect is more enhanced for $k=2$ where \texttt{kPathO} is severely restricted in number of edges stored, showing an improvement of $50-80\%$ as compared to $33-50\%$ for higher $k$. For \texttt{kLevN} again, the exceptional performance for higher values of $k$ is attributed to the broomstick structure and optimizing space usage. Notice that the impact of changing $k$ is not evident in \texttt{kLevO} as compared to \texttt{kLevN}, which exploits the available edges well. Even for $k=2$, the performance of \texttt{kLevN} is exceptional, showing an improvement of $\approx 50\%$ despite the already good performance of \texttt{kLevO}. For higher $k$, this improvement increases to $50-75\%$.
%  for all values of k take an almost constant number of passes after a certain number of edges, this is because of the broomstick structure~\cite{BaswanaGK18} of the DFS tree of random uniform graphs.

% \texttt{kLevN}(2) goes from 1 to 2 passes for around 2000 to 3000 edges i.e., $n(k-1)$.

In \Cref{fig:randVarK}, we evaluate the number of passes required for $n=1000$ by varying $k$ for different edge densities of the graph, namely (1) $O(n\log n)$, (2) $O(n\sqrt{n})$ and (3) $O(n^2)$. As expected, both \texttt{kPathN} and \texttt{kLevN} require fewer passes as the value of $k$ increases, and it improves by at least one pass over the previous due to \texttt{H1} heuristic. For \texttt{kPathO} and \texttt{kPathN} we can observe a staircase pattern for the variation of $k$ around $m=(n-1)k$ for $m=O(n\log n),O(n\sqrt{n})$, however the decrease for $m=O(n^2)$ is gradual. This can be attributed to the fact that the DFS tree for $m=O(n^2)$ is a single path; hence, with an increase in $k$, the probability of finding the path in the first pass increases. For smaller values of $k$, \texttt{kPathN} performs extremely well, reducing passes by $80-90\%$ as to larger values where improvement ranges around $33-50\%$. These can be attributed to \texttt{H3} heuristic, where the algorithm efficiently uses $nk$ space. One of the key observations is that \texttt{kLevO} is almost constant for increasing $k$, whereas \texttt{kLevN} reduces from $2$ to $1$ at small values of $k=5$ reaching optimal. Also, for the graphs of different densities, the expected steps from 2 to 1 were at different $k$ similar to \texttt{kPathN}. However, it reaches optimal way earlier because of \texttt{H3} exploiting the broomstick property~\cite{BaswanaGK18}. We thus see an improvement of $33-75\%$ for smaller $k$ but optimal performance with an improvement of $66-75\%$ for larger $k$.

\begin{observation}
On evaluation over random graphs, we have:
\begin{enumerate}
    \item \texttt{kLevN}  
    outperforms all algorithms in almost all cases. \item \texttt{kPathN} uses optimal single pass when $m\leq n(k-1)$, while \texttt{kLevN} much earlier.
    \item \texttt{kPathN} improves \texttt{kPathO} by $\approx 80\%$ for smaller $k$ and by $33-50\%$ for larger $k$.
    \item  \texttt{kLevN} improves \texttt{kLevO} by $\approx 50\%$ for smaller $k$ while up to $75\%$ for larger $k$. 
    \item The strongest impact for \texttt{kLevN} is apparently from \texttt{H3} exploiting the broomstick property.
\end{enumerate}
\end{observation}

\subsection{Random Power Law Graphs}

\begin{figure}[!h]
  \begin{subfigure}{.5\textwidth}
  \centering    \includegraphics[trim={0 1cm 0 0},clip,width=\linewidth]{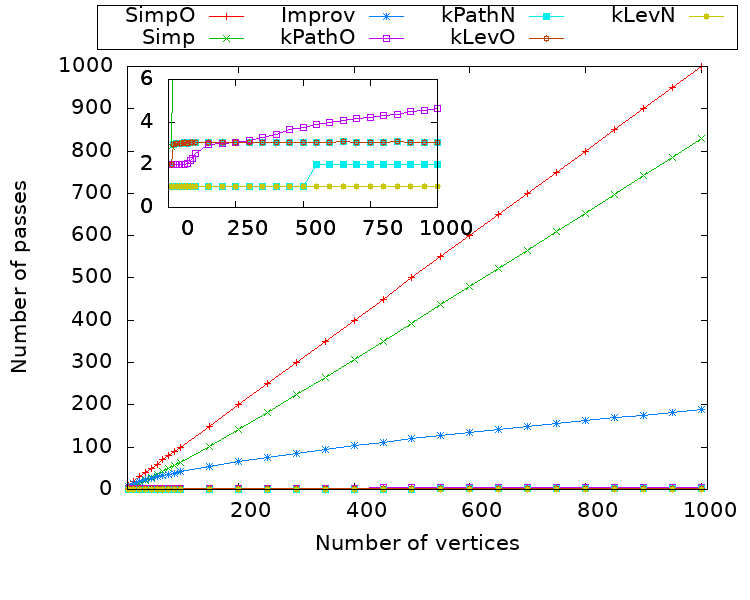}
    \caption{Density $m=O(n\log n)$}
  \end{subfigure}%
  \begin{subfigure}{.5\textwidth}
  \centering
    \includegraphics[trim={0 1cm 0 0},clip,width=\linewidth]{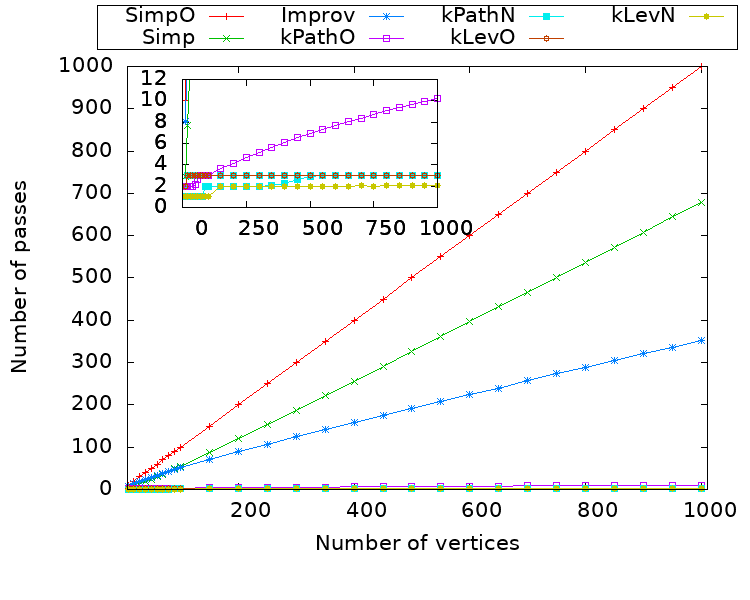}
    \caption{Density $m=O(n\sqrt{n})$}
  \end{subfigure}
  \caption{
Performance of the faster algorithms as the number of vertices is varied with (a) $m=O(n\log n)$, and (b) $m=O(n\sqrt{n})$ densities, using $10n$ edges ($k=10$).}
  \label{fig:powlawVarN}
\end{figure}

\begin{figure}[!h]
  \begin{subfigure}{.5\textwidth}
  \centering
    \includegraphics[trim={0 1cm 0 0},clip,width=\linewidth]{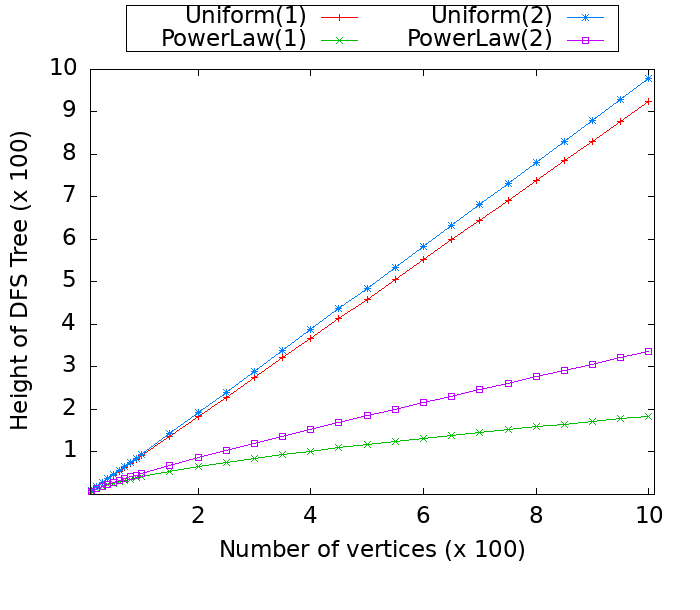}
    \caption{Variation in the average height of DFS tree.}
    \label{fig:DFSheightUvP}
  \end{subfigure}
  \begin{subfigure}{.5\textwidth}
  \centering
    \includegraphics[trim={0 1cm 0 0},clip,width=\linewidth]{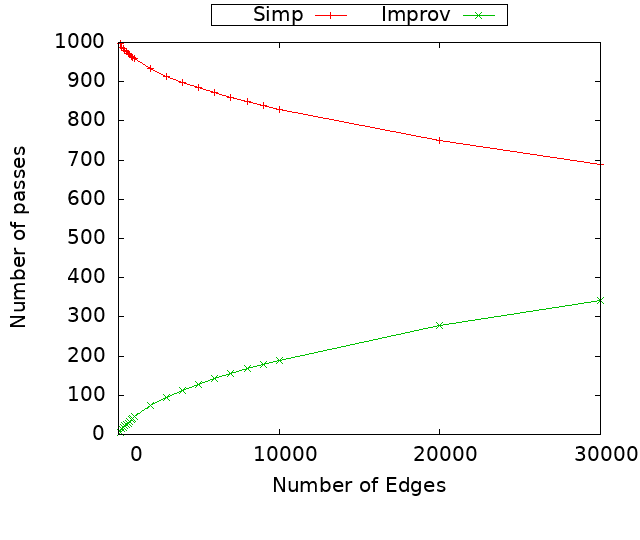}
    \caption{\texttt{Simp} and \texttt{Imprv} on varying edges.}
    \label{fig:simpImprvVarM}
  \end{subfigure}%
  \caption{Evaluation of \texttt{Simp} and \texttt{Imprv}. (a) Variation in heights of DFS trees with the number of vertices for (1) $m=O(n\log n)$ and (2) $m=O(n\sqrt{n})$ edges. (b) Performance of \texttt{Simp} and \texttt{Imprv} as the number of edges are varied up to $O(n\sqrt{n})$ for $n=1000$ vertices.}
\end{figure}

\begin{figure}[!h]
  \begin{subfigure}{.5\textwidth}
  \centering
    \includegraphics[trim={0 1cm 0 0},clip,width=\linewidth]{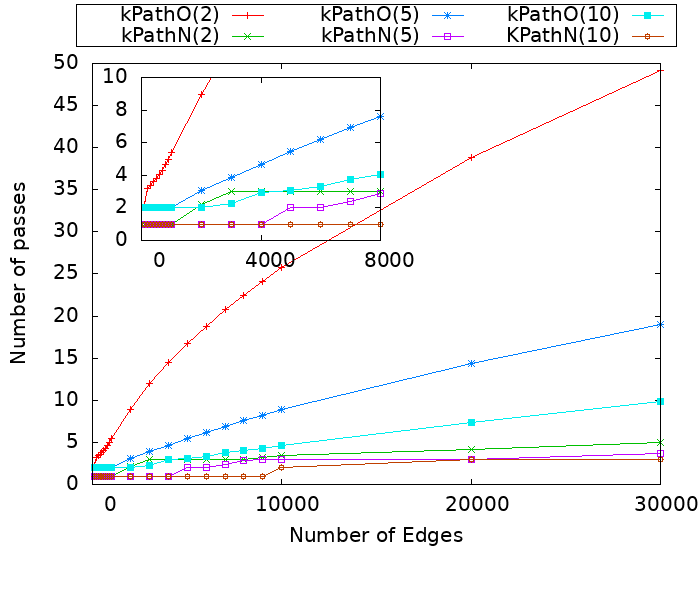}
    \caption{Improvement in \texttt{kPath}}
  \end{subfigure}%
  \begin{subfigure}{.5\textwidth}
  \centering
    \includegraphics[trim={0 1cm 0 0},clip,width=\linewidth]{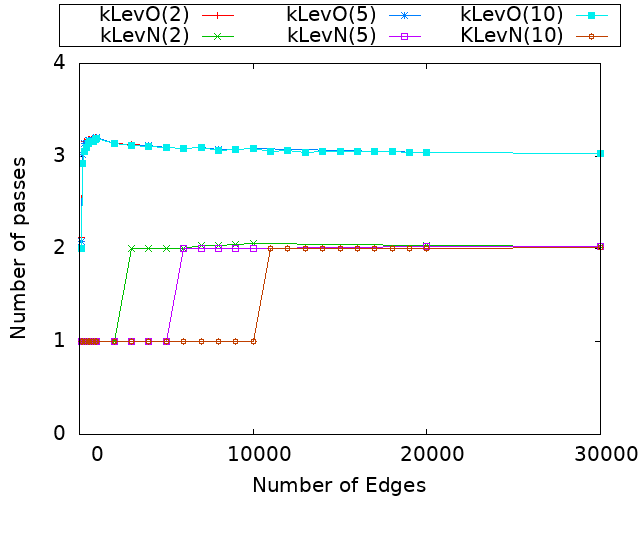}
    \caption{Improvement in \texttt{kLev}}
  \end{subfigure}
  \caption{Performance of the algorithms with variation in edges up to $m=O(n\sqrt{n})$ for $n=1000$, having different values of allowed edges ($k=2,5,10$) for (a) \texttt{kPath}, and (b) \texttt{kLev}.} % VarM for n = 1000 and m$\_$max = nC2 for K = 5,10 }
    \label{fig:powlawVarM}
\end{figure}

\begin{figure}[!h]
  \begin{subfigure}{.5\textwidth}
  \centering
    \includegraphics[trim={0 1cm 0 0},clip,width=\linewidth]{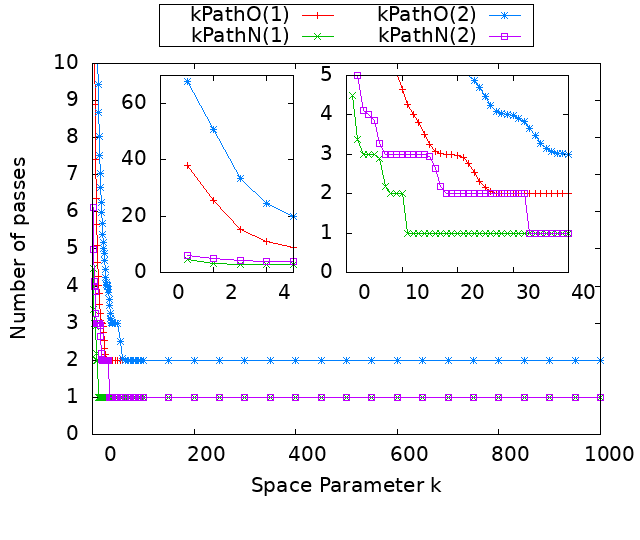}
    \caption{Improvement in \texttt{kPath}}
  \end{subfigure}%
  \begin{subfigure}{.5\textwidth}
  \centering
    \includegraphics[trim={0 1cm 0 0},clip,width=\linewidth]{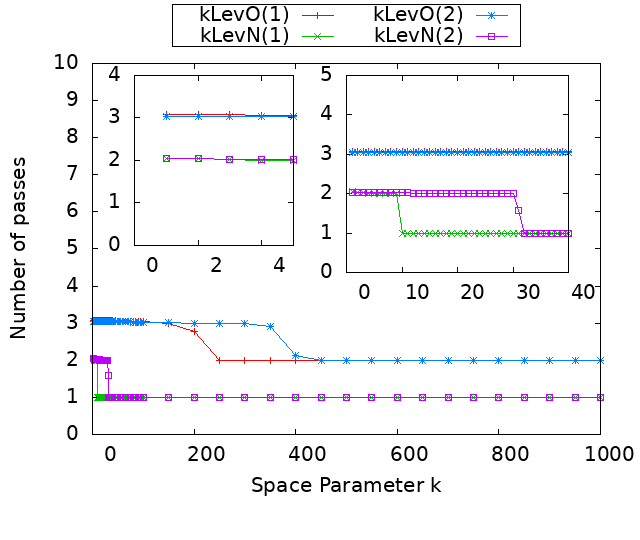}
    \caption{Improvement in \texttt{kLev}}
  \end{subfigure}
 \caption{Performance of algorithms as the number of stored edges are varied up to $k=n$ for $n=1000$ vertices and different graph densities $m$ ($O(n\log n), O(n\sqrt{n})$ for (a) \texttt{kPath}, and (b)  \texttt{kLev}.} %VarK for n = 1000 and (1) = nlogn (2) = n$\sqrt{n}$ (3) = nC2}
\label{fig:powlawVarK}
\end{figure}

In random power law graphs, we similarly perform three kinds of experiments evaluating the performance of the algorithms by varying the three parameters: the number of vertices $n$ (\Cref{fig:powlawVarN}), the number of edges $m$ 
(\Cref{fig:simpImprvVarM} and \Cref{fig:powlawVarM}), and allowed stored edges parameter $k$ (\Cref{fig:powlawVarK}). Since Khan and Mehta~\cite{KhanM19} did not evaluate the algorithms on Power law graphs, we additionally include the simpler algorithms (recall \Cref{sec:prevWork}) in the evaluation for the sake of comparison.

From \Cref{fig:powlawVarN}, we can see that, unlike random uniform graphs (see \Cref{fig:corrected-random} in \Cref{apn:prevC}), \texttt{Improv} is performing better than \texttt{Simp} in random power law graphs, this is because, in power law graphs, there are a large number of vertices with a low degree, reducing the height of the overall tree benefiting \texttt{Improv}. This difference reduces for $m =O(n\sqrt{n})$ as compared to $m=O(n\log n)$ due to the apparent increase in the height of the DFS tree, which in turn aids \texttt{Simp} as more vertices can be added in a single pass. We verify our inference by plotting the height of the DFS trees for uniformly random and power law graphs for the two densities in \Cref{fig:DFSheightUvP}. The variation clearly resembles our expectation, explaining the performance of \texttt{Simp} and \texttt{Imprv}.  The improvement of \texttt{kPathN} and \texttt{kLevN} is again at least a pass, while both again perform exceptionally well, improving around $50-75\%$ and $50-66\%$ respectively for $m=O(n\log n)$. For higher density, this improvement increases for \texttt{kPathN} to $\approx 80\%$ and decreases for \texttt{kLev} to $33\%$.

% \begin{figure}[!h]
%   \centering
%     \includegraphics[width=.5\linewidth]{arxiv_images/powlaw/new/VarM_1000_SP3_Avg.png}
%   \caption{Performance of the simple algorithms as the number of edges are varied upto $O(n\sqrt{n})$ for $n=1000$ vertices . } 
%     \label{fig:powlawsimpimprovVarM}
% \end{figure}

 \Cref{fig:simpImprvVarM,fig:powlawVarM} shows the performance of all the algorithms for $n=1000$ as the number of edges is increased to $O(n\sqrt{n})$. The performance of \texttt{Simp} and \texttt{Imprv} are again as expected according to variation in the height of the DFS tree.
\texttt{kPathN} and \texttt{kLevN} again improves \texttt{kPathO} and \texttt{kLevO} by a huge margin, where improvement in \texttt{kPath} reduces with density ($60-90\%$) and in  \texttt{kLev} increases with density ($33-66\%$). Again, the improvement in \texttt{kPath} is greater for smaller $k$ ($80-90\%$), while that of \texttt{kLev} is greater for larger $k$ ($66\%$). For both \texttt{kPathN} and \texttt{kLevN}, we see a step from 1 to 2 passes when the number of edges crosses $(n-1)k$ as expected. The impact of increasing $k$ is also clearly visible in \texttt{kLevN} as compared to \texttt{kLevO}. Note that the step for \texttt{kPathO} from 2 to 3 is earlier than $nk$ edges, which may be because of the duplicate edges due to which extra passes are required. Also, the number of passes increases with the number of edges, which is not similar to the case of random uniform graphs. This is because of no particular structural property of DFS trees for power law graphs as compared to the broomstick property. However, despite the huge improvement of \texttt{kPathN} over \texttt{kPathO} as compared to \texttt{kLevN} over \texttt{kLevO}, we still have the state-of-the-art performance by \texttt{kLevN}.

In \Cref{fig:powlawVarK}, we again observe that steps from 2 to 1 are formed for \texttt{kLevN} and \texttt{kPathN} when the number of edges is around $nk$, due to the impact of \texttt{H2} heuristic. 
%initially $k'$ is set to infinity and it gets modified only after the number of edges reaches $nk$ if the number of edges is less than $nk$ then $k'$ remains infinity which makes sure that vertices of all levels are added to DFS tree in the first pass itself. 
Note that for random uniform graphs, this step occurs earlier than $nk$ edges because of the broomstick property~\cite{BaswanaGK18}, which is leveraged by the \texttt{H3} heuristic. 
The improvement in \texttt{kPath} is up to $90\%$ for smaller values of $k$ and at least $50\%$ for larger values. On the other hand, the improvement in \texttt{kLev} ranges from $33-66\%$ for initial values and then reduces to $50\%$.

\begin{observation}
On evaluation over random power law graphs, we have:
\begin{enumerate}
    \item \texttt{Improv} performs better than \texttt{Simp} (opposite to the case of random uniform graphs), where difference decreases with increasing edge density due to the height of the DFS tree.
    \item \texttt{kPathN} improves \texttt{kPathO} from $50-90\%$, which is higher ($80-90\%$) for smaller $k$.% reducing the worst case from $70$ to $5$.
    \item \texttt{kLevN} improves \texttt{kLevO} from $33-66\%$, which is higher ($50-66\%$) for larger $k$. %having worst case reduced from $3$ to $2$.
    \item Both \texttt{kPathN} and \texttt{kLevN} require  \textbf{optimal} one pass if $m\leq n(k-1)$.
    \item The number of passes required in the worst-case reduced from $3$ to \textbf{merely} $2$ for \texttt{kLev}, and \textbf{significantly reduced} from $70$ to $5$ for \texttt{kPath}.
%    \item In the variation of n plots, the difference between \texttt{Simp} and \texttt{Improv} is less in the case of graphs with m = $O(n\sqrt{n})$ compared to m = $O(n\log n)$.
\end{enumerate}
\end{observation}

\section{Conclusion}
We performed an experimental analysis of the practical semi-streaming algorithms for computing a DFS tree. We presented additional heuristics to improve the \texttt{kPath} and \texttt{kLev} algorithms. In real graphs, the performance improved by $45-90\%$ for \texttt{kPath} and  $40-70\%$ for \texttt{kLev}  on average based on the value of $k$, requiring optimal one pass in $\approx 40\%$ and $\approx 50\%$ cases. 
In random graphs as well, the heuristics improves \texttt{kPath} from $30-90\%$, which is greater for smaller $k$, and \texttt{kLev} from $33-75\%$ which is greater for larger $k$. Additionally, we frequently have optimal one pass and significantly improved worst-case performance of \texttt{kPath} from 70 to 5 passes and of \texttt{kLev} from 3 to merely 2 passes for around 1000 vertices.

%In random graphs as well, the heuristics results frequently in optimal one pass, typically improving around $50-100\%$.

Overall, prior to this study, the state-of-the-art algorithm was \texttt{kLev}, which computed the DFS tree in a few passes independent of $k$, never reaching optimal. However, the algorithm \texttt{kPath}, which is arguably simpler to understand and implement, performed poorly for smaller values of $k$ and is almost similar to \texttt{kLev} as $k$ is increased. Firstly, the heuristics always show a significant improvement and achieve the optimal one-pass computation for the majority of test cases (which was zero earlier). Secondly, the improved \texttt{kLev} is able to exploit greater space, improving the performance more significantly for higher values of $k$. Thirdly, the improved \texttt{kPath} now performs exceptionally well for all values of $k$, showing more significant improvement for smaller values. Thus, in case the difference of a few passes is insignificant, we recommend using the simpler \texttt{kPath} while the state-of-the-art algorithm is still \texttt{kLev}, which is relatively complex.

\begin{comment}
<<<for conclusions>> 
In this paper, we delved into the experimental evaluation of Depth-First Search (DFS) in the semi-streaming model and identified the shortcomings in existing state-of-the-art methods and propose heuristics to enhance their performance for Kpath(n/k) and Klev(h/k) algorithms. These modified algorithms have shown exception improvements in experimental analysis. There is an approx. 39.5% and 60.2% improvements for Kpath and Klev algorithms respectively in case of real graphs  and significant improvements in the analysis done for random graphs. We have successfully managed to reduce the number of passes for most datasets down to 1 and 2.
\end{comment}

\bibliography{paper}

\appendix

\section{Corrected results from Khan and Mehta~\cite{KhanM19}}
\label{apn:prevC}

\begin{table}

    \centering
    \scalebox{0.8}{
    \begin{tabular}{ |p{1.5cm}|p{1cm}|p{1cm}|p{1cm}| p{1cm}|p{1cm}|p{1cm}|p{0.6cm}|p{0.6cm}|p{0.6cm}| p{1cm}|p{0.6cm}|p{0.6cm}|p{0.6cm}|   }
    \hline
    Dataset & n  & m & m/n & Simp & Imprv & \texttt{kPath} - n & 2n & 5n & 10n & \texttt{kLev} - n & 2n & 5n & 10n \\
    \hline
     CU   & 49    & 107 &   2.18 & 23 & 32&  \fbox{5} & 4& 2&  2 & 3 & 3 & 3 & 3\\[0.1cm]
     AJazz   & 198    & 2.74K &   13.85 & 53&154& \fbox{19} & \fbox{16}& 5& 3& 3& 3& 3& 3\\[0.1cm]
     HM   & 2.43K    & 16.6K &   6.86 &1.31K&753& \fbox{47} & 13& \fbox{5}& 2& 4& 4& 4& 4\\[0.1cm]
     ArxAP   & 18.8K    & 198K &   10.55 & 9.36K& 6.49K & \fbox{231} & \fbox{33}& \fbox{10}& \fbox{3}& 5& 5& 5& 5\\[0.1cm]
     AsCaida   & 26.5K    & 53.4K &   2.02 & 24.7K& 979 & \fbox{43} & 8& 2& 2& 4& 4& 4& 4\\[0.1cm]
     BrightK   & 58.2K    & 214K &   3.68 &  43.3K &10.3K & \fbox{242} & \fbox{14}& 2& 2& 5& 5& 5& 5\\[0.1cm]
     LMocha   & 104K    & 2.19M &   21.07& 66.4K & 40K&  \fbox{850} & 22& 6& \fbox{3}& 4& 4& 4& 4\\[0.1cm]
     FlickrE   & 106K    & 2.32M &   21.87& 55K &51.7K& \fbox{737} & \fbox{43}& \fbox{7}& 4& 5& 5& 5& 5\\[0.1cm]
     WordNet   & 146K    & 657K &   4.50 & 96.9K & 23.7K& \fbox{303} & \fbox{66}& \fbox{4}& 2& 6& 6& 6& 6\\[0.1cm]
     Douban   & 155K    & 327K &   2.11 &  145K & 11.5K & \fbox{284}& 7& 2& 2& 4& 4& 4& 4\\[0.1cm]
     Gowalla   & 197K    & 950K &   4.83 & 134K & 45.3K & \fbox{617} & \fbox{32}& \fbox{6}& 2& 6& 6& 6& 6\\[0.1cm]
     Dblp   & 317K    & 1.05M &   3.31 &  214K & 42.3K & \fbox{471} & \fbox{43}& 2& 2& 6& 6& 6& 6 \\[0.1cm]
     Amazon   & 355K    & 926K &   2.76 &  204K & 80.1K & \fbox{309} & \fbox{107}& 2& 2& 6& 6& 6& 6 \\[0.1cm]
     \hline
    \end{tabular}}
    \caption{Corrected version of \texttt{kPath} and \texttt{kLev} where corrections are hightlighted.}
    \label{tab:reduce-pass}
\end{table}

\begin{figure}[h]
  \begin{subfigure}{.5\textwidth}
  \centering
    \includegraphics[trim={0 1.3cm 0 0},clip,width=.98\linewidth]{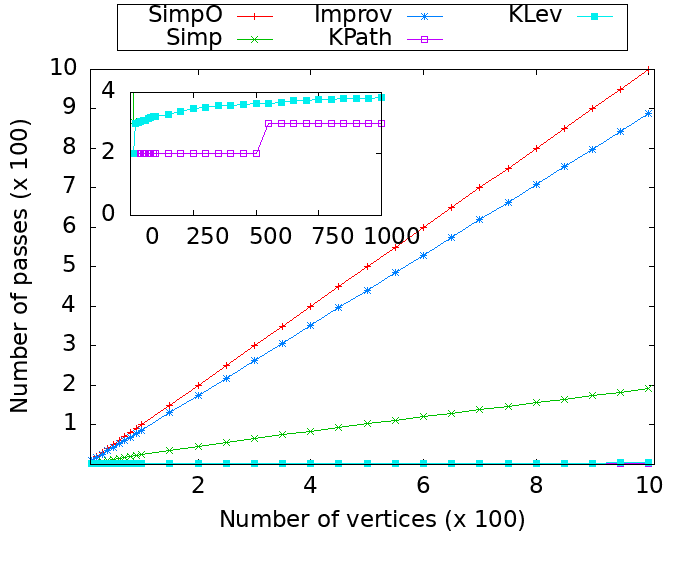}
    % \caption{VarN for n = 1000 and m = nlogn(left) or nrootn(right)}
     \caption{Number of vertices varied}
  \end{subfigure}%
  \begin{subfigure}{.5\textwidth}
  \centering
    \includegraphics[trim={0 1.3cm 0 0},clip,width=.98\linewidth]{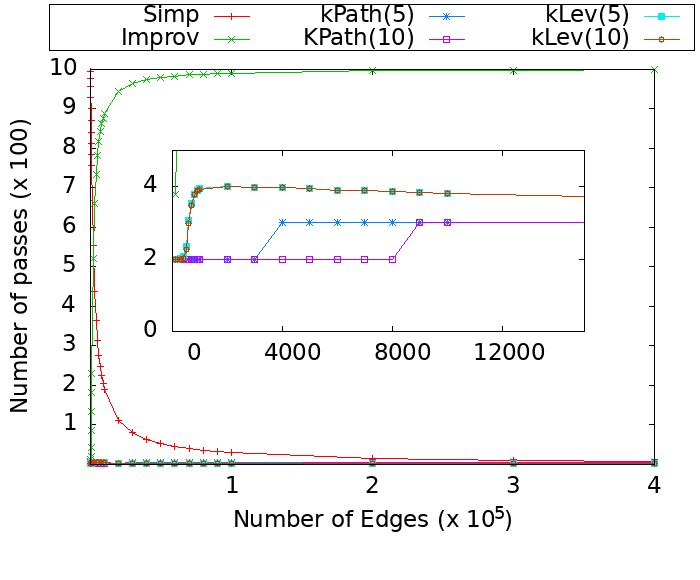}
    % \caption{VarM for n = 1000 and m$\_$max = nC2}
    \caption{Number of edges varied}
  \end{subfigure}
\centering
  \begin{subfigure}{.5\textwidth}
  \centering
    \includegraphics[trim={0 1.3cm 0 0},clip,width=.98\linewidth]{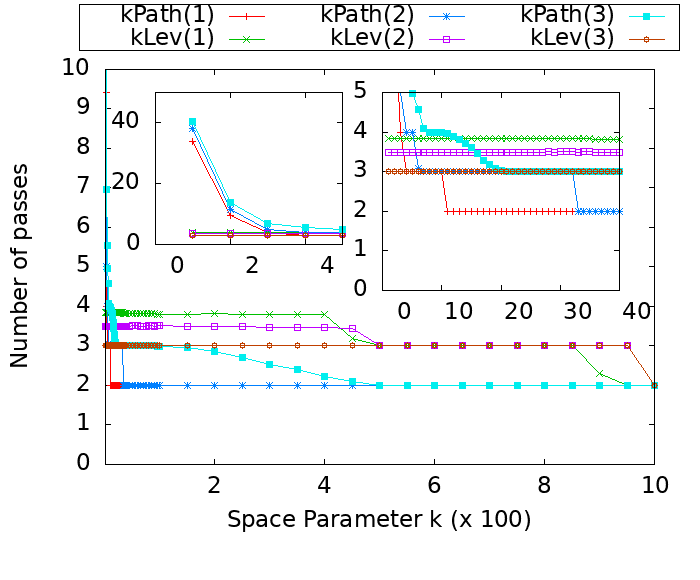}
    % \caption{VarK for n = 1000 and (1) = nlogn (2) = nrootn (3) = nC2}
    \caption{Number of Stored edges varied}
  \end{subfigure}
  \caption{Corrected Versions of performance of algorithms in \cite{KhanM19}. (a) Number of passes required by different algorithms when $m=O(n\log n)$ and $k=10$, (b) Number of passes required by different algorithms when $n=1000$ and $k=5,10$, and (c) Number of passes required by different algorithms when $n=1000$ and $m$ is (1) $O(n\log n)$, (2) $O(n\sqrt{n})$ and (3) $O(n^2)$. }
  \label{fig:corrected-random}
\end{figure}

The algorithm \texttt{kPath} stores a spanning tree for each component having $n'$ vertices and stores additional $n'k$ edges. The DFS tree on the combined set of edges is computed, and the longest path is added. However, this clearly violates the $nk$ space-bound as edges of the spanning trees are not considered. We thus report the corrected pseudocode in \Cref{apn:pseudocodes} and the updated results of the previous paper in \Cref{tab:reduce-pass} and \Cref{fig:corrected-random}.

%Table [\ref{tab:reduce-pass}] corresponds to analysis of corrected versions of KPath and KLev Algorithms on Real Graphs. Figure (\ref{fig:corrected-random}) corresponds to analysis of corrected versions of KPath and KLev Algorithms on Random Graphs.

%We have corrected the above observations and did an experimental analysis on real graphs. It is observed that incase of KPath algorithm, there is significant change in no of passes,however incase of KLev Algorithm, there is no change in no of passes as compared to previous analysis in paper[4] as explained above. We have added the corrected version of analysis table[5] on real graphs for KPath and KLev Algorithms in appendix[6].

%\multirow{3}{*}{2.18}

\begin{procedure}
%\BlankLine
$E'_C\leftarrow \emptyset$\;
\tcc{Initiate a pass over $E$, for all components in parallel}
\While(\tcc*[f]{process first $|V_C|(k-1)$ edges of $C$})
{\textcolor{blue}{$|E'_C|\leq |V_C|(k-1)$} or Stream is over}
{
%\tcc{Let $e$ be the next edge in input stream}
\lIf{next edge $e$ belongs to $C$}
{$E'_C\leftarrow E'_C\cup \{e\}$}
}

\BlankLine
\lIf{$par(r_C)\neq \phi$}{Add $(r_C,par(r_C))$ to $T$}
$T'_C\leftarrow$ DFS tree of $T_C\cup E'_C$ from root $r_C$\;
%\BlankLine
\lIf(\tcc*[f]{pass over $E$ was completed}){\textcolor{blue}{$|E'_C|\cup T_C\leq |V_C|\cdot k$}}
{Add $T'_C$ to $T$}

\Else{
$P\leftarrow$ Path from $r_C$ to lowest vertex in $T'_C$\;
Add $P$ to $T$\;
\BlankLine
\tcc{Continue the pass for all components in parallel}
\ForAll{edges in $E'_C$ followed by the remaining pass over $E$}
{
%\ForAll{edges in a single pass over $E$}{
Compute the components $C_1,...,C_f$ of $C\setminus P$ using Union-Find algorithm\;
\tcc{This essentially computes $T_{C_1},...,T_{C_f}$}
Find lowest edge $e_i$ from each component $C_i$ to $P$\;
}

\ForEach{Component $C_i$ of $C\setminus P$}{
%Add $e_i$ to $T$\;
$par(y_i)\leftarrow$ $x_i$
\tcc*[r]{Let $e_i=(x_i,y_i)$, where $y_i\in C_i$}
\ref{alg:KPathO}($C_i$,$T_{C_i}$,$y_i$)\;
}

}
\caption{Compute-DFS-kPathO($C$,$T_C$,$r_C$): Computes a DFS tree of the 
%subgraph of $G$ induced by the 
component $C$ rooted at the vertex $r_C\in C$.}
\label{alg:KPathO}
\end{procedure}	

\begin{procedure}
\BlankLine
${\cal E}_C\leftarrow \{e\}$\;
\While{${{\cal E}_C}\neq \emptyset$}
{
	
	$(x,y)\leftarrow$ Extract an edge from ${{\cal E}_C}$
	\tcc*{where $level(x)\geq level(y)$}
	$w\leftarrow $ LCA of $x$ and $y$ in $T_C$\;
	\BlankLine
	\If(\tcc*[f]{$(x,y)$ is a cross edge}){$w\neq x$ and $w\neq y$}
		{
		$v\leftarrow $ Ancestor of $y$ whose parent in $T_C$ is $w$\;
		Remove the edge $(w,v)$ from $T_C$\;
		Reverse the parent-child relationship of edges on path from $y$ to $v$ in $T_C$\;
		Add the edge $(x,y)$ to $T_C$\;
		
		$E_R\leftarrow$ The cross edges from $H_C\setminus {\cal E}_C$ in the tree $T_C$\;
		${{\cal E}_C}\leftarrow {{\cal E}_C}\cup E_R$\;
		}	
%	Add $(x,y)$ to $H_C$\;
}
\caption{Maintain-DFS($T_C$,$e$): Maintains the DFS tree $T_C$ on insertion an edge 
$e$ abiding {\em monotonic fall} making $T_C$ a valid DFS tree of $H_C$.}
\label{alg:rebuild}
\end{procedure}	

\begin{procedure}
%\BlankLine
Initialize $H_C\leftarrow T_C$\; %, Compute $rep[v],\forall v\in V_C$\;

\tcc{Initiate a pass over $E$, for all components in parallel}
\ForEach(\tcc*[f]{edges in the input stream from $C$})
{edge $(x,y)$ in $E$ if $(x,y)\in E_C$}
{
\lIf{$x$ and $y$ are within the same tree in $T_C\setminus T'_C$}{Continue}
%\lIf(\tcc*[f]{ignore edge of lower component}){$rep[x]= rep[y]$}{Continue}
		{
		Add $(x,y)$ to $H_C$\;
%		\BlankLine
%		\If{$(x,y)$ is a cross edge}
%			{
			\ref{alg:rebuild}($T_C,(x,y)$)\;
%			\ref{alg:rebuild}($x,y,T_C$)\;
			Remove excess edges from $H_C$
			\tcc*[r]{non-tree edges in $T_C\setminus T'_C$}
%			Update $H_C$ and $rep[v],\forall v\in V_C$\;
%			}	
		}
}

$T'_{C}\leftarrow$ Top $k$ levels of $T_C$
\tcc*[r]{vertices $v$ with $0\leq level\leq k-1$}
			\BlankLine
\lIf{$par(r_C)\neq \phi$}{Add $(r_C,par(r_C))$ to $T$}
Add $T'_{C}$ to $T$\;

			\BlankLine

\ForEach{tree $\tau \in T_C\setminus T'_C$}
{
$v \leftarrow root(\tau)$\;
\tcc{Let $C_v$ be component containing $v$ in $C \setminus T'_C$}
$par(v)\leftarrow$ Parent of $v$ in $T_C$\;
\ref{alg:KLevO}($C_v$,$T_C(v)$,$v$):
}

\caption{Compute-DFS-kLevO($C$,$T_C$,$r_C$): Computes a DFS tree of the 
component $C$ whose spanning tree $T_C$ is rooted at the vertex $r_C\in C$.}
\label{alg:KLevO}
\end{procedure}

\pagebreak

\section{Detailed results of evaluation on real datasets}
\label{apn:detailReal}

% From Table [\ref{tab:KPathold_KPathnew_RedPer}] we can see that for k=1 the percentage of improvement is less comparatively. This is because the number of passes required is already high for k=1, so although the improvement in the number of passes is high the percentage is relatively less. For greater values of k, in many cases even though the improvement in the number of passes is by 1 pass the percentage of improvement is very high as the number of passes are very low. For \texttt{kPath}, in very few cases the number of passes required is increased which we can observe in the case of ArxAP dataset. 

\begin{figure}[h]
  \begin{subfigure}{.33\linewidth}
  \centering
    \includegraphics[trim={0 2cm 0 0},clip,width=.98\linewidth]{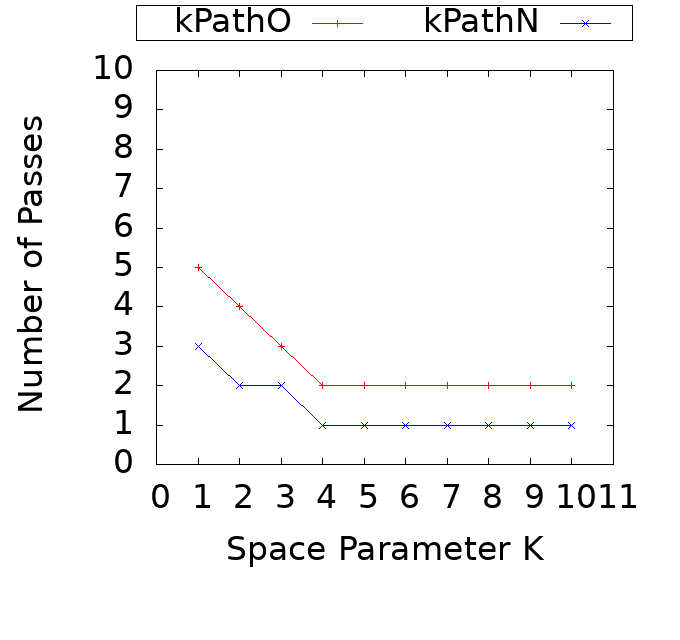}
    \caption{CU}
  \end{subfigure}%
  \begin{subfigure}{.33\linewidth}
  \centering
    \includegraphics[trim={0 2cm 0 0},clip,width=.98\linewidth]{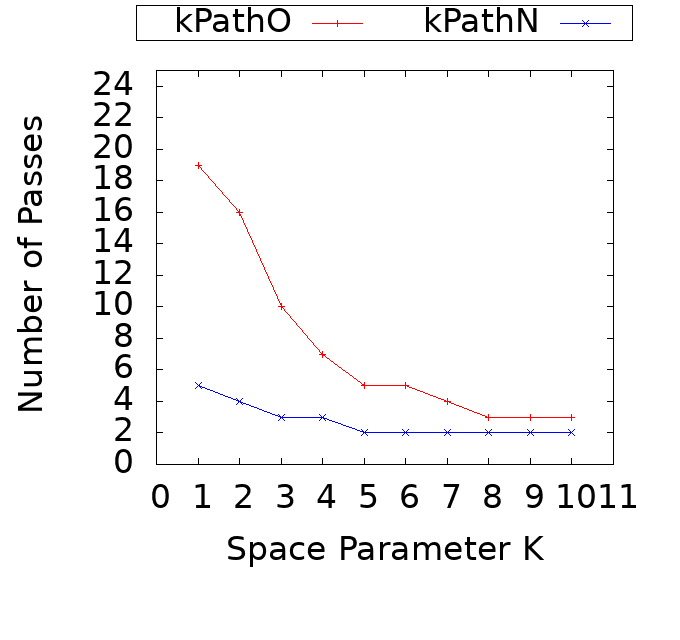}
    \caption{AJazz}
  \end{subfigure}
  \begin{subfigure}{.33\linewidth}
  \centering
    \includegraphics[trim={0 2cm 0 0},clip,width=.98\linewidth]{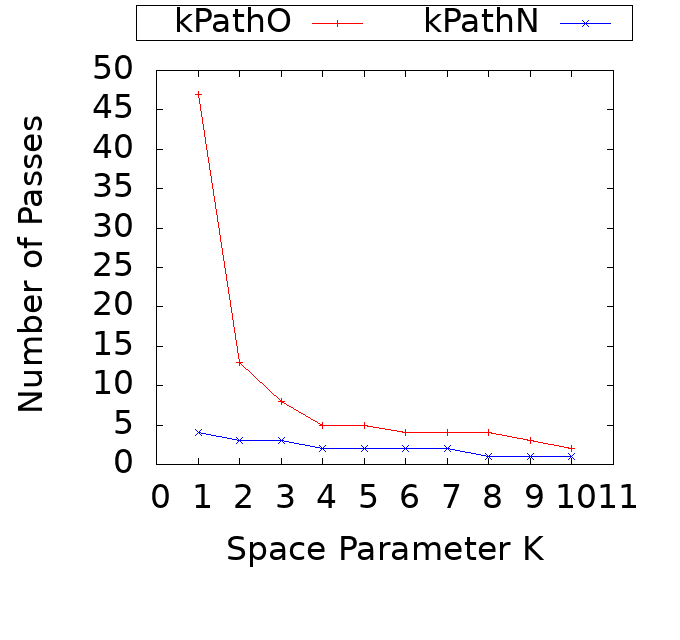}
    \caption{HM}
  \end{subfigure}%
  \newline
  \begin{subfigure}{.33\textwidth}
  \centering
    \includegraphics[trim={0 2cm 0 0},clip,width=.98\linewidth]{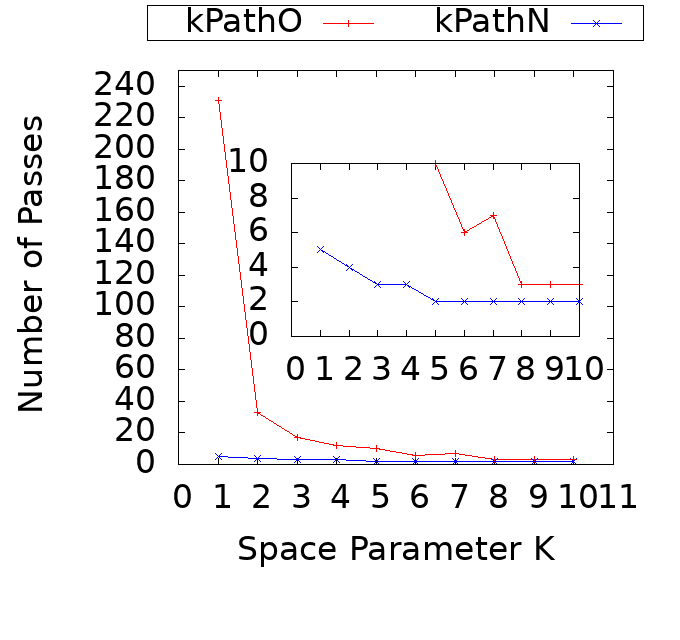}
    \caption{ArxAP}
  \end{subfigure}
  \begin{subfigure}{.33\textwidth}
  \centering
    \includegraphics[trim={0 2cm 0 0},clip,width=.98\linewidth]{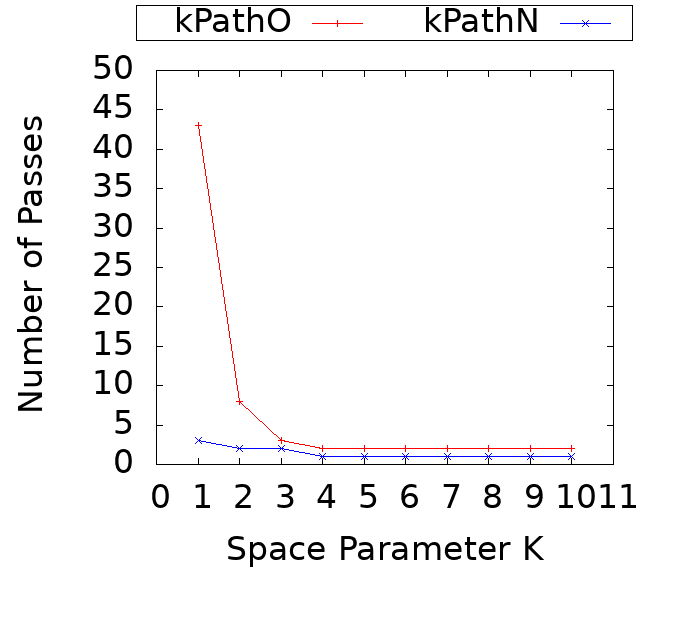}
    \caption{AsCaida}
  \end{subfigure}%
  \begin{subfigure}{.33\textwidth}
  \centering
    \includegraphics[trim={0 2cm 0 0},clip,width=.98\linewidth]{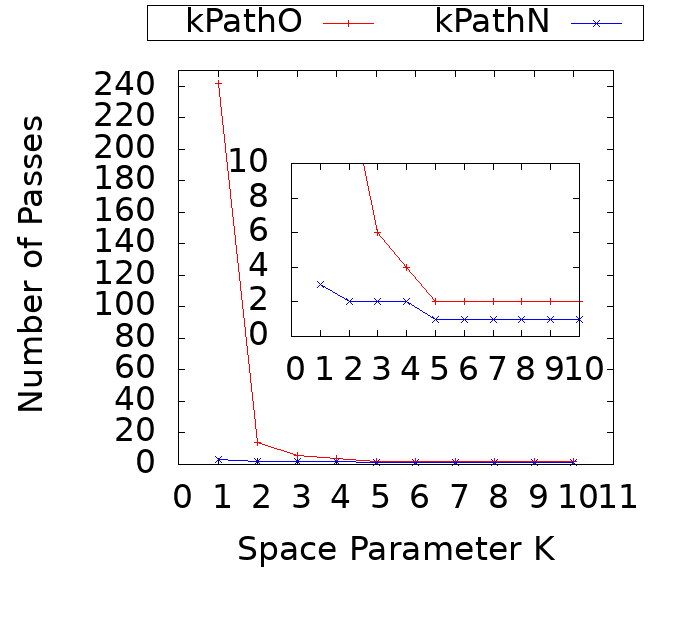}
    \caption{BrightK}
  \end{subfigure}
  \begin{subfigure}{.33\textwidth}
  \centering
    \includegraphics[trim={0 2cm 0 0},clip,width=.98\linewidth]{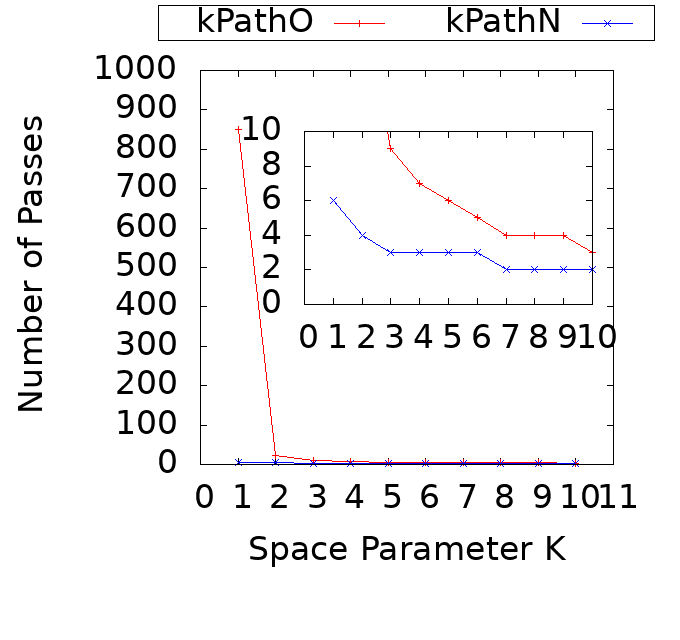}
    \caption{LMocha}
  \end{subfigure}%
  \begin{subfigure}{.33\textwidth}
  \centering
    \includegraphics[trim={0 2cm 0 0},clip,width=.98\linewidth]{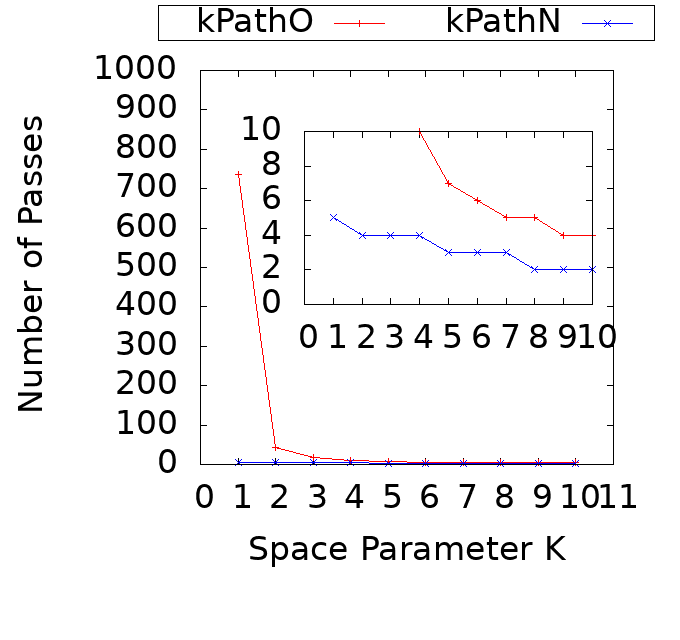}
    \caption{FlickrE}
  \end{subfigure}
  \begin{subfigure}{.33\textwidth}
  \centering
    \includegraphics[trim={0 2cm 0 0},clip,width=.98\linewidth]{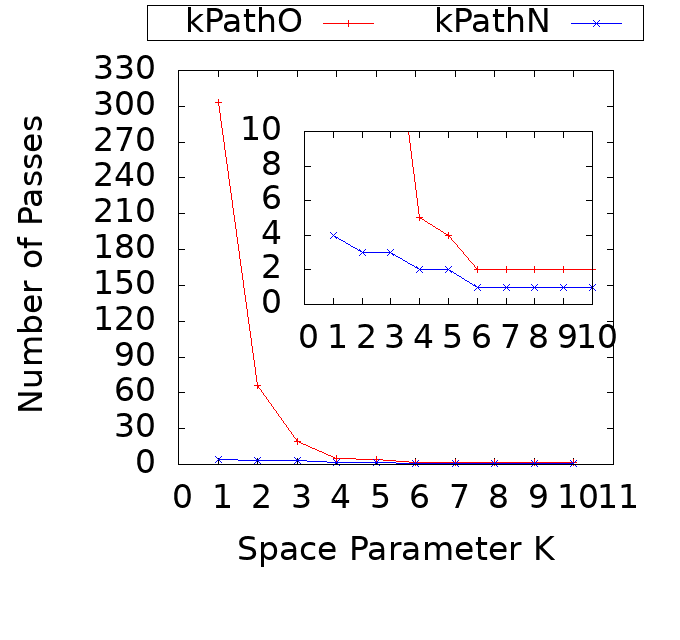}
    \caption{WordNet}
  \end{subfigure}%
  \newline
  \begin{subfigure}{.33\textwidth}
  \centering
    \includegraphics[trim={0 2cm 0 0},clip,width=.98\linewidth]{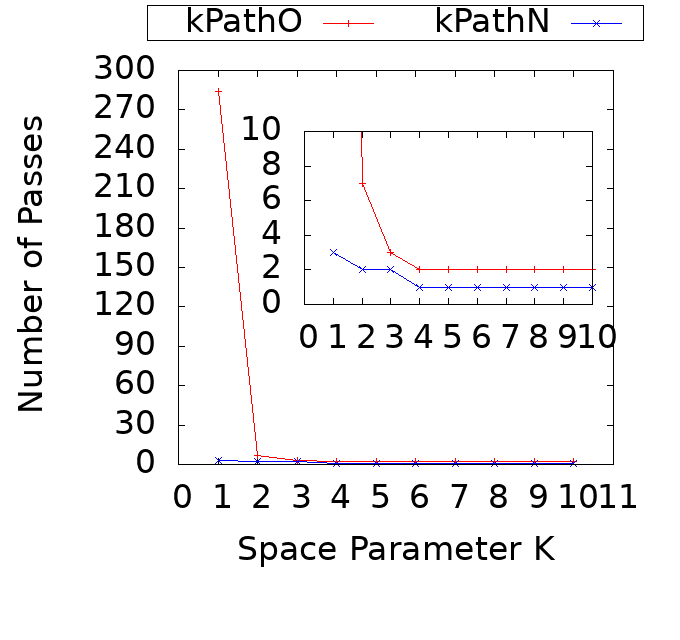}
    \caption{Douban}
  \end{subfigure}
  \begin{subfigure}{.33\textwidth}
  \centering
    \includegraphics[trim={0 2cm 0 0},clip,width=.98\linewidth]{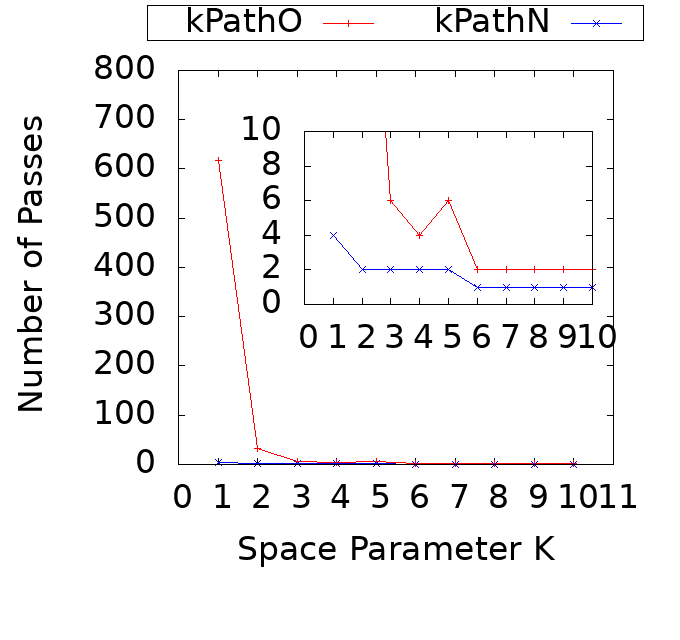}
    \caption{Gowalla}
  \end{subfigure}%
  \begin{subfigure}{.33\textwidth}
  \centering
    \includegraphics[trim={0 2cm 0 0},clip,width=.98\linewidth]{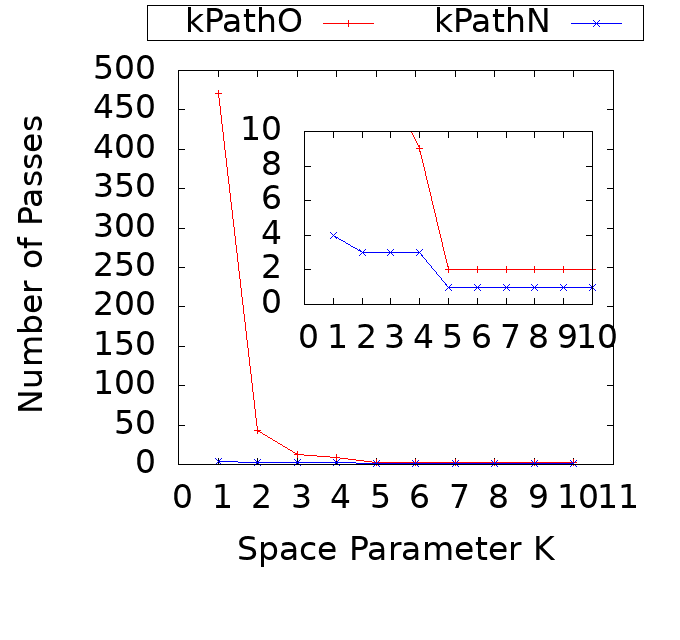}
    \caption{Dblp}
  \end{subfigure}
  \begin{subfigure}{.33\textwidth}
  \centering
    \includegraphics[trim={0 2cm 0 0},clip,width=.98\linewidth]{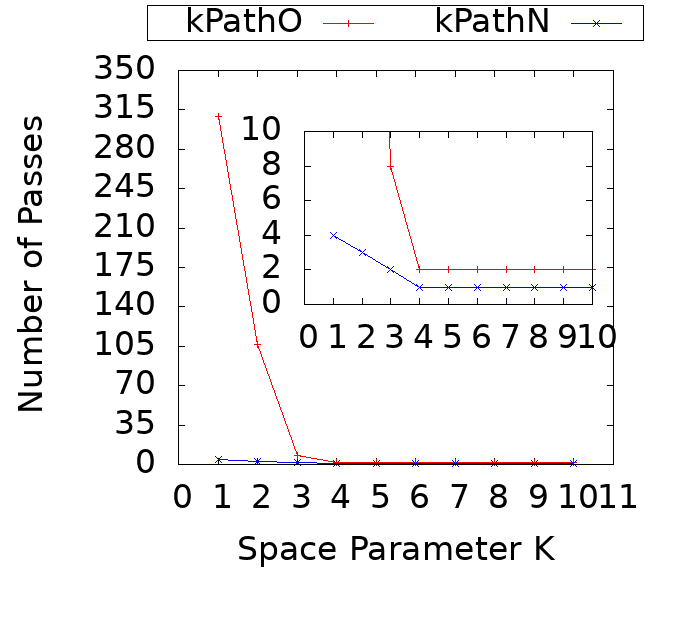}
    \caption{Amazon}
  \end{subfigure}
  \caption{Improvement in \texttt{kPath} Algorithm for Real Graphs}
  \label{fig:kpath-new-real}
\end{figure}

\begin{figure}[h]
  \begin{subfigure}{.33\textwidth}
  \centering
    \includegraphics[trim={0 2cm 0 0},clip,width=.98\linewidth]{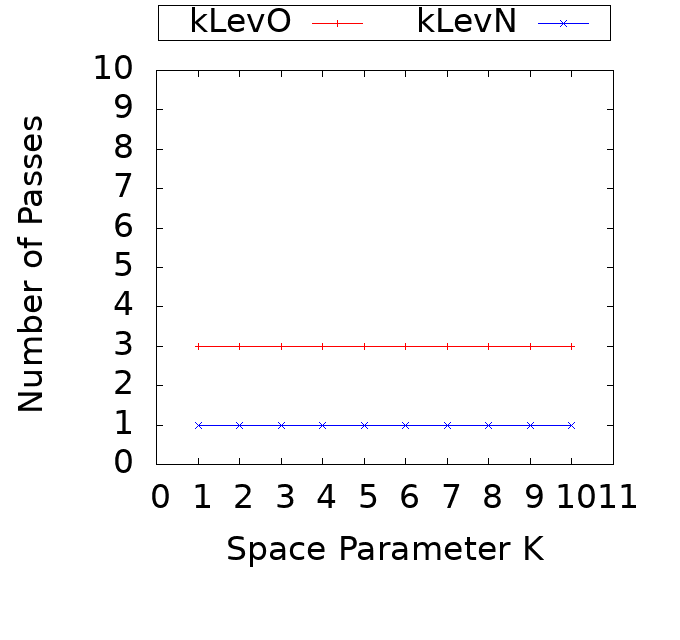}
    \caption{CU}
  \end{subfigure}%
  \begin{subfigure}{.33\textwidth}
  \centering
    \includegraphics[trim={0 2cm 0 0},clip,width=.98\linewidth]{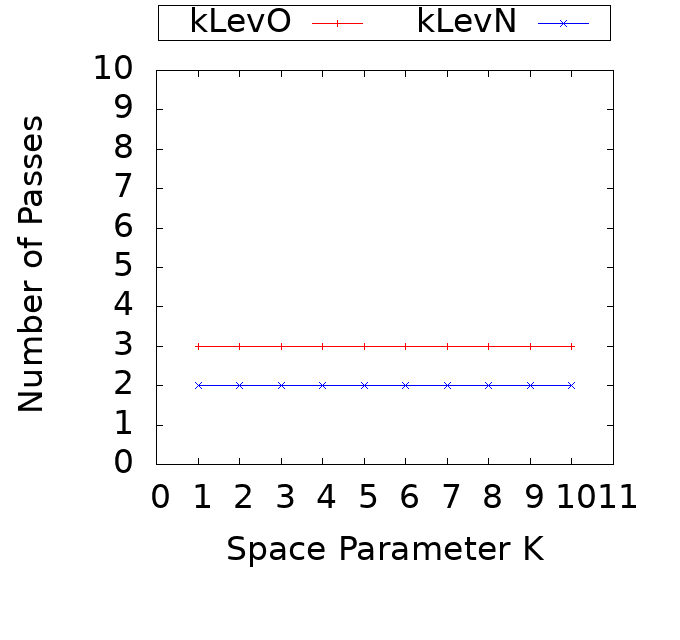}
    \caption{AJazz}
  \end{subfigure}
  \begin{subfigure}{.33\textwidth}
  \centering
    \includegraphics[trim={0 2cm 0 0},clip,width=.98\linewidth]{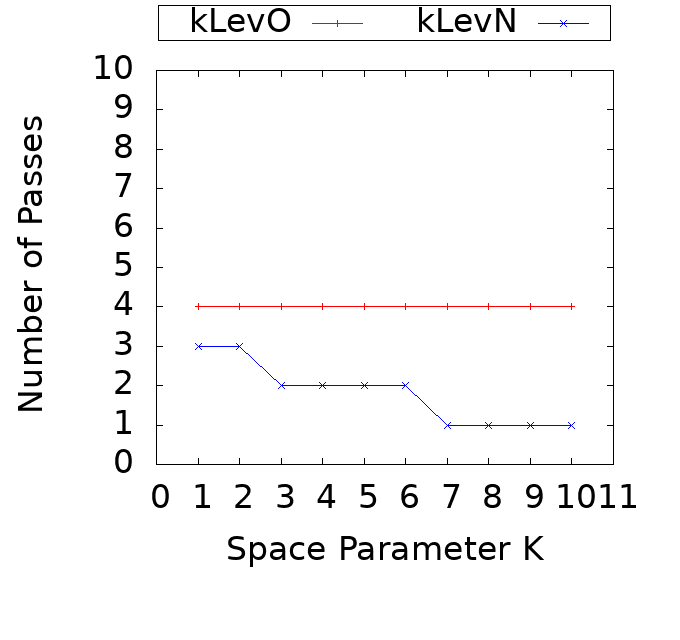}
    \caption{HM}
  \end{subfigure}%
  \newline
  \begin{subfigure}{.33\textwidth}
  \centering
    \includegraphics[trim={0 2cm 0 0},clip,width=.98\linewidth]{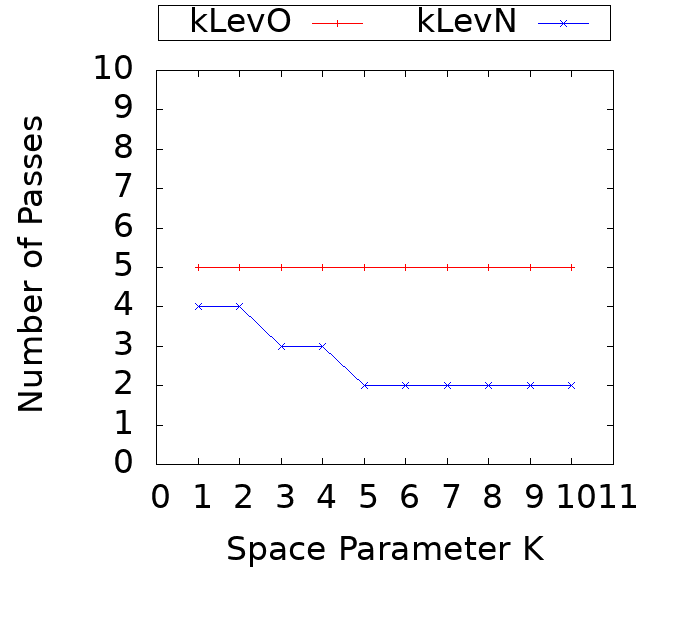}
    \caption{ArxAP}
  \end{subfigure}
  \begin{subfigure}{.33\textwidth}
  \centering
    \includegraphics[trim={0 2cm 0 0},clip,width=.98\linewidth]{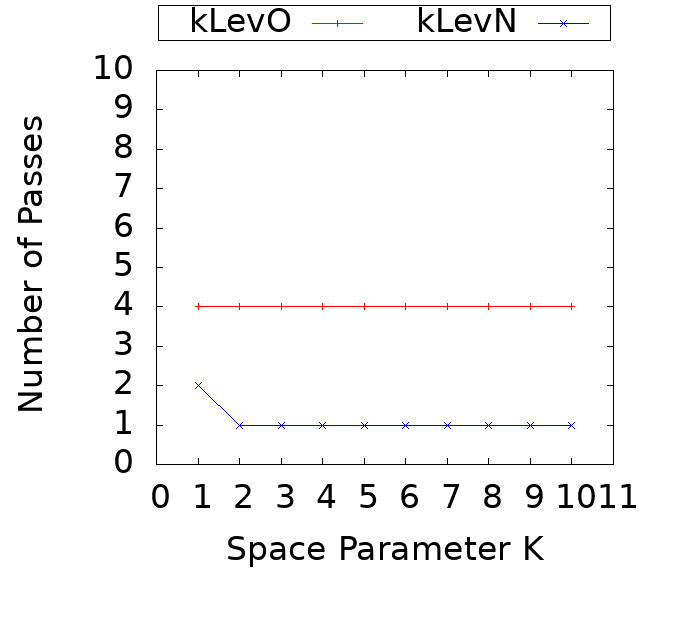}
    \caption{AsCaida}
  \end{subfigure}%
  \begin{subfigure}{.33\textwidth}
  \centering
    \includegraphics[trim={0 2cm 0 0},clip,width=.98\linewidth]{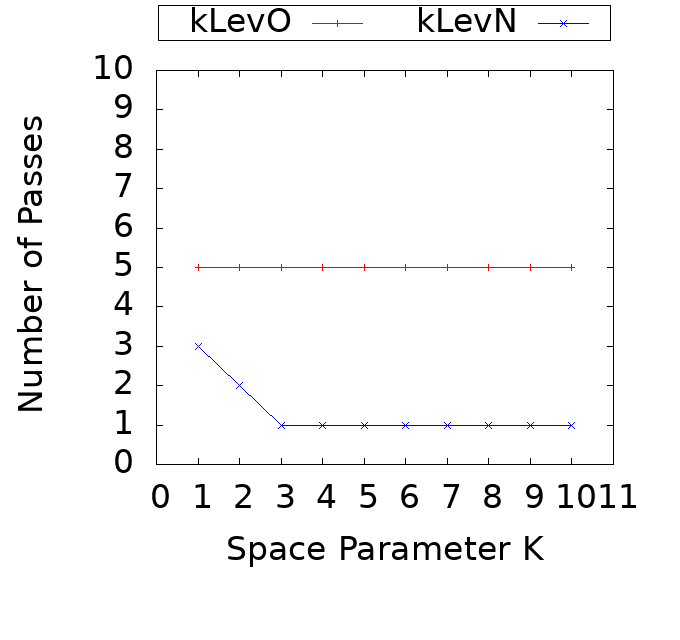}
    \caption{BrightK}
  \end{subfigure}
  \newline
  \begin{subfigure}{.33\textwidth}
  \centering
    \includegraphics[trim={0 2cm 0 0},clip,width=.98\linewidth]{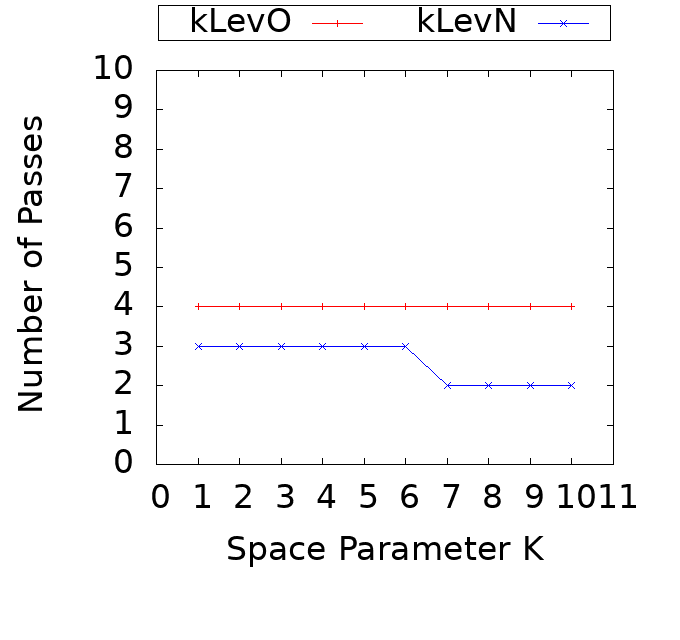}
    \caption{LMocha}
  \end{subfigure}%
  \begin{subfigure}{.33\textwidth}
  \centering
    \includegraphics[trim={0 2cm 0 0},clip,width=.98\linewidth]{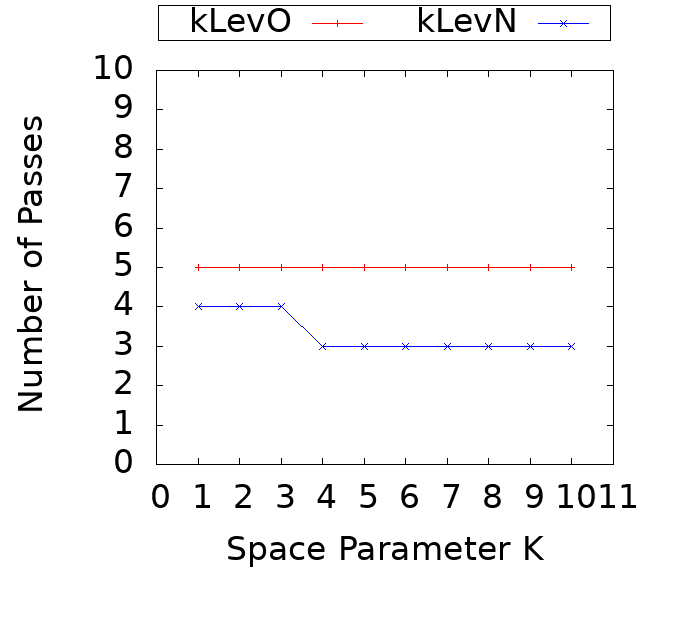}
    \caption{FlickrE}
  \end{subfigure}
  \begin{subfigure}{.33\textwidth}
  \centering
    \includegraphics[trim={0 2cm 0 0},clip,width=.98\linewidth]{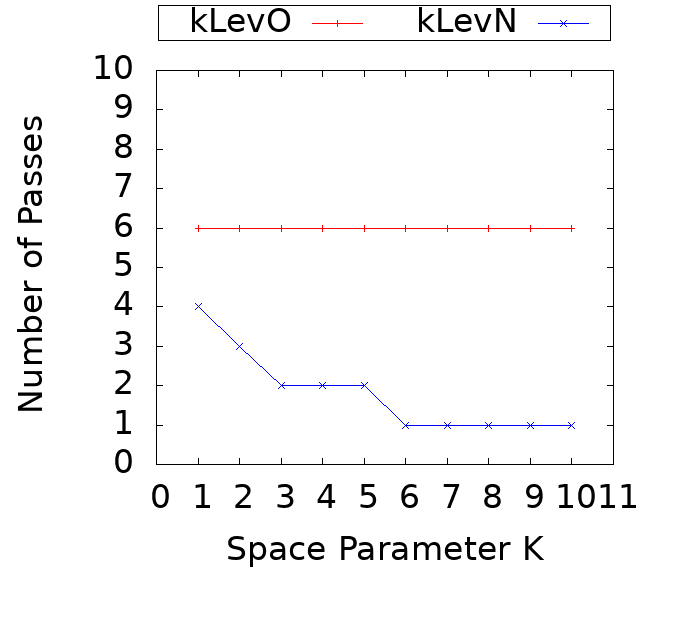}
    \caption{WordNet}
  \end{subfigure}%
  \newline
  \begin{subfigure}{.33\textwidth}
  \centering
    \includegraphics[trim={0 2cm 0 0},clip,width=.98\linewidth]{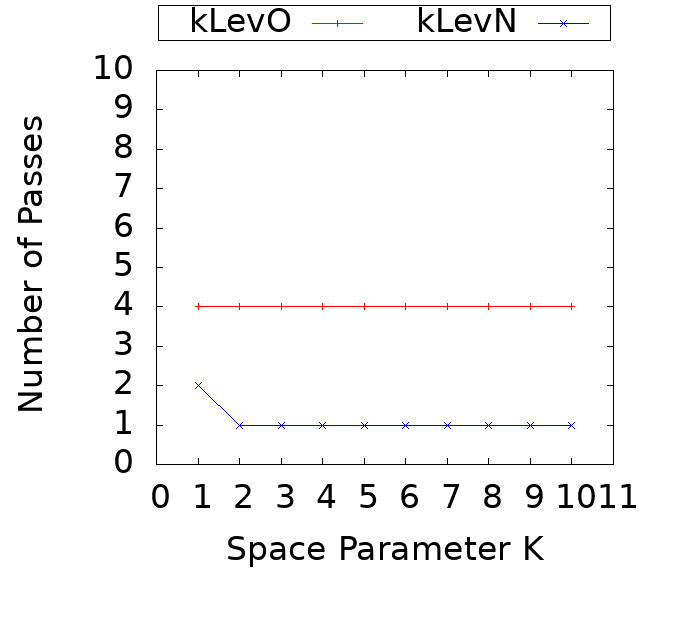}
    \caption{Douban}
  \end{subfigure}
  \begin{subfigure}{.33\textwidth}
  \centering
    \includegraphics[trim={0 2cm 0 0},clip,width=.98\linewidth]{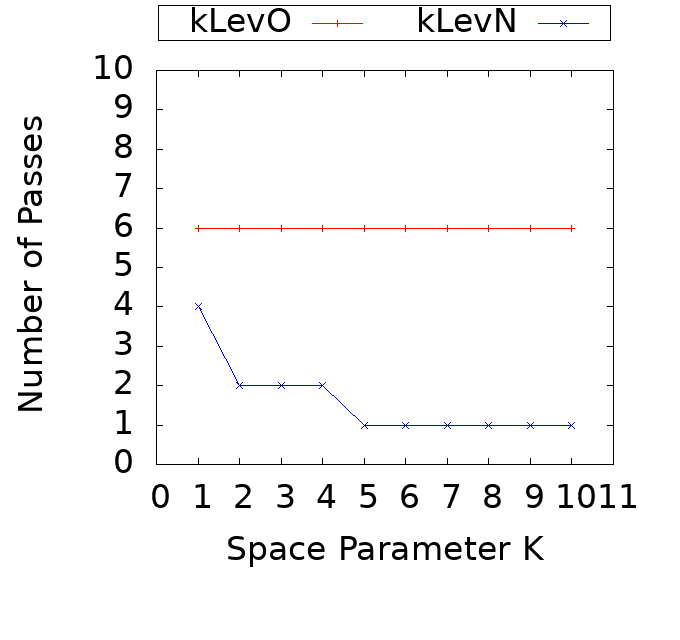}
    \caption{Gowalla}
  \end{subfigure}%
  \begin{subfigure}{.33\textwidth}
  \centering
    \includegraphics[trim={0 2cm 0 0},clip,width=.98\linewidth]{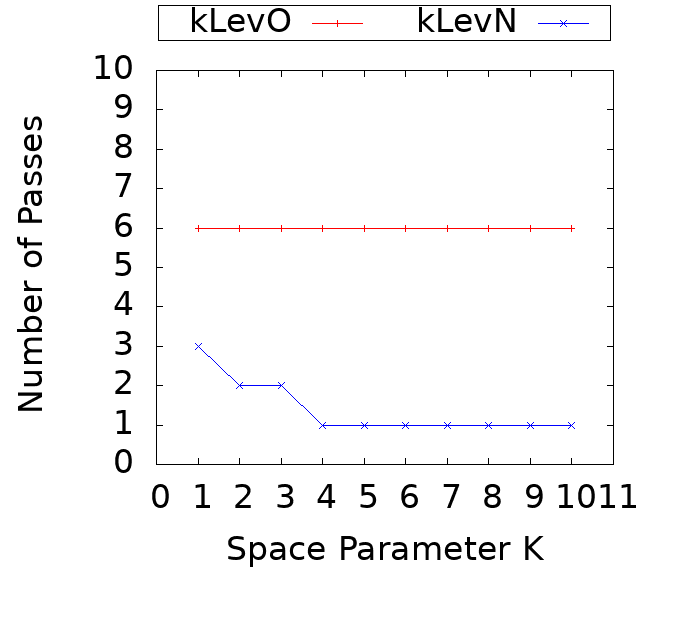}
    \caption{Dblp}
  \end{subfigure}
  \begin{subfigure}{.33\textwidth}
  \centering
    \includegraphics[trim={0 2cm 0 0},clip,width=.98\linewidth]{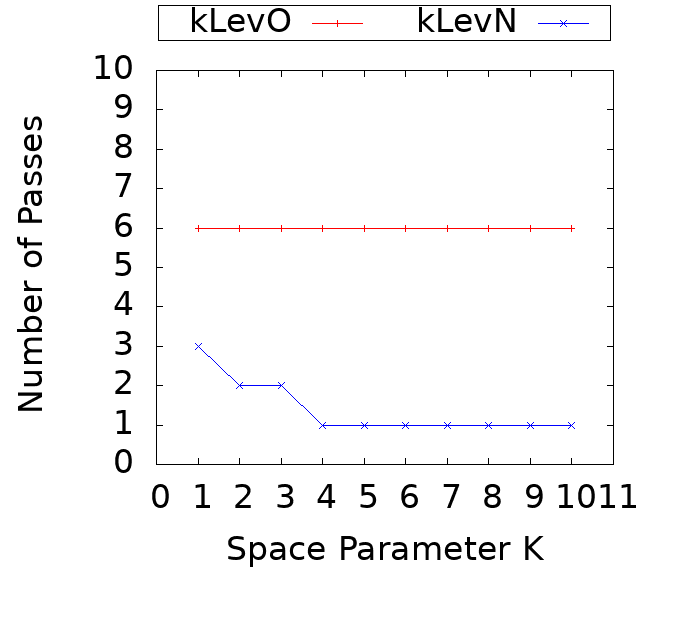}
    \caption{Amazon}
  \end{subfigure}
   \caption{Improvement in \texttt{kLev} Algorithm for Real Graphs}
  \label{fig:klev-new-real}
\end{figure}

\begin{figure}[h!]
  \begin{subfigure}{1\columnwidth}
  \centering
    \includegraphics[trim={0 1.8cm 0 0},clip,width=1.1\columnwidth]{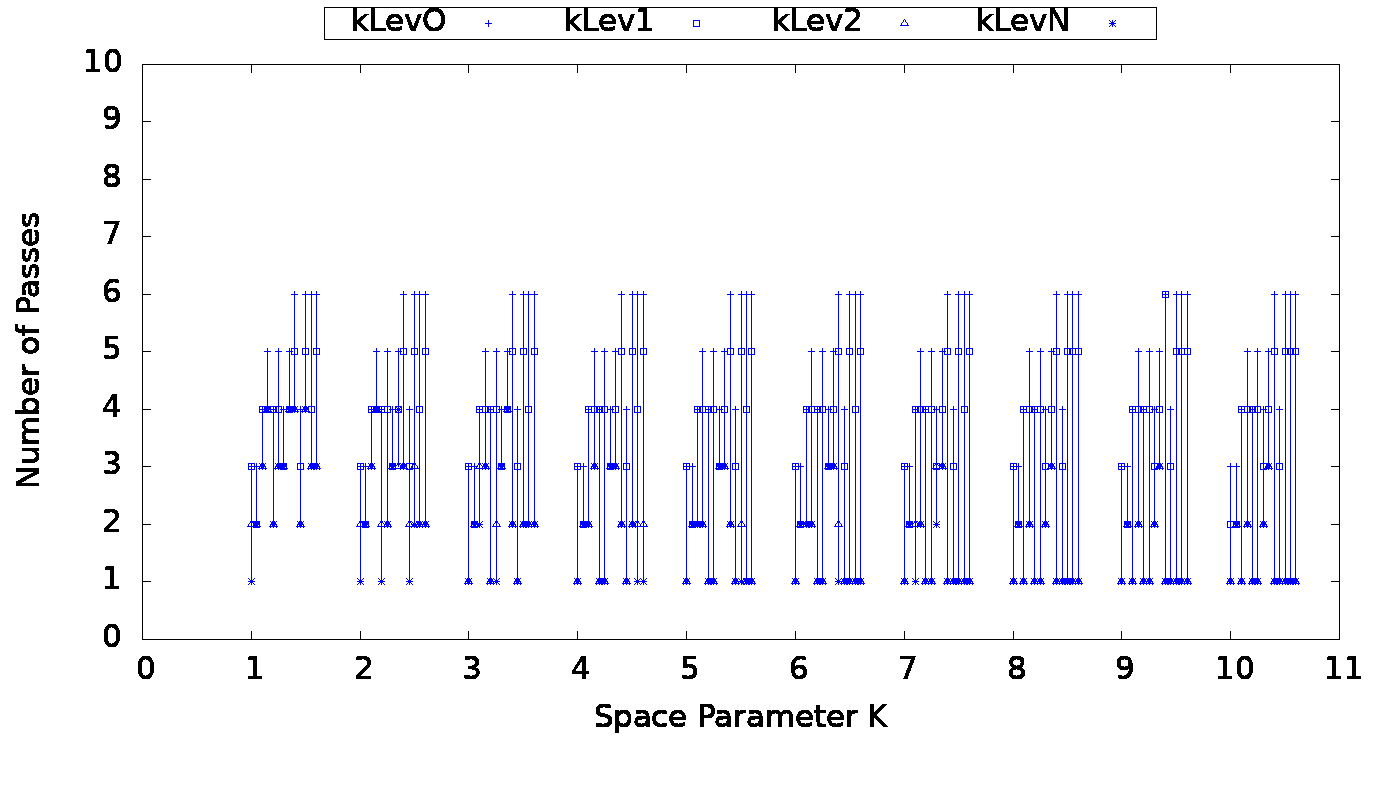}
  \end{subfigure}
  \caption{Improvement in required passes for \texttt{kLev} for real datasets .}
  \label{fig:realindividualImprovkLev}
\end{figure}

\begin{figure}[h!]
  \begin{subfigure}{1\columnwidth}
  \centering
    \includegraphics[trim={0 1.8cm 0 0},clip,width=1.1\columnwidth]{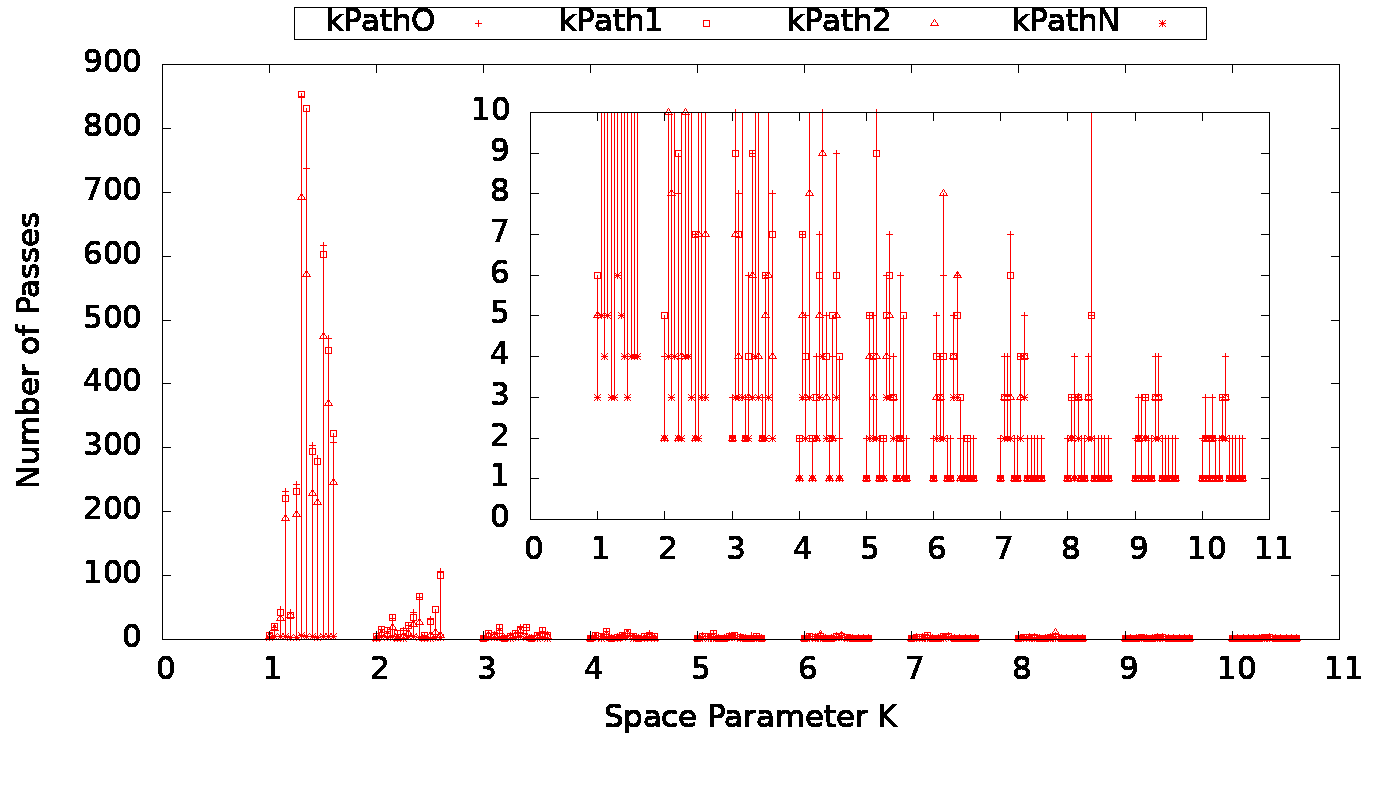}
  \end{subfigure}
  \caption{Improvement in required passes for \texttt{kPath}  for real datasets .}
  \label{fig:realindividualImprovkPath}
\end{figure}

From \Cref{tab:KPathold_KPathnew_RedPer}, we can see that the percentage of improvements for \texttt{kPath} are drastic. Especially for smaller values of $k$, the percentage of improvement is above $90\%$ and decreases with an increase in the value of $k$. The main reason behind this is that the previous algorithm is just utilizing $n^*k$ space per pass, which is low for smaller values of $k$ compared to larger values of $k$. As we proceed through the algorithm, the number of unvisited vertices decreases, thus decreasing the stored edges. By introducing \texttt{H2} (optimal $nk$ edge space) heuristic, we try to utilize the complete $nk$ space in every pass. For \texttt{kLev} from \Cref{tab:KLevold_KLevnew_RedPer}, we can see that in most cases, the number of passes is reduced to $1$. For larger values of $k$, our heuristics show a greater impact on the results of \texttt{kLev}, which is expected because, as the available space increases, we can efficiently use the $nk$ space and add more levels from the total $nk$ space utilization heuristic. The same data is plotted in \Cref{fig:kpath-new-real} and \Cref{fig:klev-new-real} showing the overall improvement in \texttt{kPath} and \texttt{kLev}. In the main paper, we showed a summarized version of improvements due to each heuristic on real datasets for different $k$. We plot the corresponding data for each dataset individually in \Cref{fig:realindividualImprovkLev} and \Cref{fig:realindividualImprovkPath}.

\begin{table}
\centering
\scalebox{0.75}{
    \begin{tabular}{ |p{1.3cm}|p{0.9cm}|p{0.9cm}|p{0.7cm} |p{1.3cm}|p{.8cm}| p{.5cm}|p{.5cm}|p{.5cm}|p{.5cm}|p{.5cm}| p{.5cm}|p{.5cm}|p{.5cm}|p{.5cm}|   }
       \hline
Dataset & $n$ & $m$ & $m/n$ & Heuristic & $k=1$ & $2$ & $3$ & $4$ & $5$ & $6$ & $7$ & $8$ & $9$ & $10$\\
\hline
\multirow{3}{*}{CU} & \multirow{3}{*}{49} & \multirow{3}{*}{107} & \multirow{3}{*}{2.18} & \texttt{kPathO} & 5 & 4 & 3 & 2 & 2 & 2 & 2 & 2 & 2 & 2\\
& & &  & \texttt{kPath1} & 6 & 5 & 2 & 2 & 1 & 1 & 1 & 1 & 1 & 1\\
& & &  & \texttt{kPath2} & 5 & 2 & 2 & 1 & 1 & 1 & 1 & 1 & 1 & 1\\
& & &  & \texttt{kPathN} & 3 & 2 & 2 & 1 & 1 & 1 & 1 & 1 & 1 & 1\\
& & &  & Red$\%$ & 40 & 50 & 33 & 50 & 50 & 50 & 50 & 50 & 50 & 50\\
\hline
\multirow{3}{*}{AJazz} & \multirow{3}{*}{198} & \multirow{3}{*}{2.74K} & \multirow{3}{*}{13.85} & \texttt{kPathO} & 19 & 16 & 10 & 7 & 5 & 5 & 4 & 3 & 3 & 3\\
& & &  & \texttt{kPath1} & 21 & 16 & 9 & 7 & 5 & 4 & 3 & 3 & 2 & 2\\
& & &  & \texttt{kPath2} & 17 & 10 & 7 & 5 & 4 & 3 & 3 & 2 & 2 & 2\\
& & &  & \texttt{kPathN} & 5 & 4 & 3 & 3 & 2 & 2 & 2 & 2 & 2 & 2\\
& & &  & Red$\%$ & 73 & 75 & 70 & 57 & 60 & 60 & 50 & 33 & 33 & 33\\
\hline
\multirow{3}{*}{HM} & \multirow{3}{*}{2.43K} & \multirow{3}{*}{16.6K} & \multirow{3}{*}{6.86} & \texttt{kPathO} & 47 & 13 & 8 & 5 & 5 & 4 & 4 & 4 & 3 & 2\\
& & &  & \texttt{kPath1} & 43 & 14 & 7 & 4 & 4 & 3 & 3 & 3 & 1 & 1\\
& & &  & \texttt{kPath2} & 33 & 8 & 4 & 3 & 3 & 3 & 3 & 1 & 1 & 1\\
& & &  & \texttt{kPathN} & 4 & 3 & 3 & 2 & 2 & 2 & 2 & 1 & 1 & 1\\
& & &  & Red$\%$ & 91 & 76 & 62 & 60 & 60 & 50 & 50 & 75 & 66 & 50\\
\hline
\multirow{3}{*}{ArxAP} & \multirow{3}{*}{18.8K} & \multirow{3}{*}{198K} & \multirow{3}{*}{10.55} & \texttt{kPathO} & 231 & 33 & 17 & 12 & 10 & 6 & 7 & 3 & 3 & 3\\
& & &  & \texttt{kPath1} & 221 & 35 & 19 & 12 & 9 & 4 & 6 & 3 & 3 & 2\\
& & &  & \texttt{kPath2} & 190 & 19 & 13 & 8 & 4 & 8 & 3 & 3 & 2 & 2\\
& & &  & \texttt{kPathN} & 5 & 4 & 3 & 3 & 2 & 2 & 2 & 2 & 2 & 2\\
& & &  & Red$\%$ & 97 & 87 & 82 & 75 & 80 & 66 & 71 & 33 & 33 & 33\\
\hline
\multirow{3}{*}{AsCaida} & \multirow{3}{*}{26.5K} & \multirow{3}{*}{53.4K} & \multirow{3}{*}{2.02} & \texttt{kPathO} & 43 & 8 & 3 & 2 & 2 & 2 & 2 & 2 & 2 & 2\\
& & &  & \texttt{kPath1} & 38 & 9 & 2 & 2 & 1 & 1 & 1 & 1 & 1 & 1\\
& & &  & \texttt{kPath2} & 37 & 2 & 2 & 1 & 1 & 1 & 1 & 1 & 1 & 1\\
& & &  & \texttt{kPathN} & 3 & 2 & 2 & 1 & 1 & 1 & 1 & 1 & 1 & 1\\
& & &  & Red$\%$ & 93 & 75 & 33 & 50 & 50 & 50 & 50 & 50 & 50 & 50\\
\hline
\multirow{3}{*}{BrightK} & \multirow{3}{*}{58.2K} & \multirow{3}{*}{214K} & \multirow{3}{*}{3.68} & \texttt{kPathO} & 242 & 14 & 6 & 4 & 2 & 2 & 2 & 2 & 2 & 2\\
& & &  & \texttt{kPath1} & 231 & 13 & 4 & 3 & 2 & 1 & 1 & 1 & 1 & 1\\
& & &  & \texttt{kPath2} & 195 & 4 & 3 & 2 & 1 & 1 & 1 & 1 & 1 & 1\\
& & &  & \texttt{kPathN} & 3 & 2 & 2 & 2 & 1 & 1 & 1 & 1 & 1 & 1\\
& & &  & Red$\%$ & 98 & 85 & 66 & 50 & 50 & 50 & 50 & 50 & 50 & 50\\
\hline
\multirow{3}{*}{LMocha} & \multirow{3}{*}{104K} & \multirow{3}{*}{2.19M} & \multirow{3}{*}{21.07} & \texttt{kPathO} & 850 & 22 & 9 & 7 & 6 & 5 & 4 & 4 & 4 & 3\\
& & &  & \texttt{kPath1} & 853 & 22 & 9 & 6 & 5 & 4 & 4 & 3 & 3 & 3\\
& & &  & \texttt{kPath2} & 692 & 10 & 6 & 5 & 4 & 4 & 3 & 3 & 3 & 2\\
& & &  & \texttt{kPathN} & 6 & 4 & 3 & 3 & 3 & 3 & 2 & 2 & 2 & 2\\
& & &  & Red$\%$ & 99 & 81 & 66 & 57 & 50 & 40 & 50 & 50 & 50 & 33\\
\hline
\multirow{3}{*}{FlickrE} & \multirow{3}{*}{106K} & \multirow{3}{*}{2.32M} & \multirow{3}{*}{21.87} & \texttt{kPathO} & 737 & 43 & 18 & 10 & 7 & 6 & 5 & 5 & 4 & 4\\
& & &  & \texttt{kPath1} & 831 & 35 & 15 & 11 & 6 & 5 & 4 & 5 & 3 & 3\\
& & &  & \texttt{kPath2} & 571 & 23 & 12 & 9 & 5 & 6 & 4 & 11 & 3 & 3\\
& & &  & \texttt{kPathN} & 5 & 4 & 4 & 4 & 3 & 3 & 3 & 2 & 2 & 2\\
& & &  & Red$\%$ & 99 & 90 & 77 & 60 & 57 & 50 & 40 & 60 & 50 & 50\\
\hline
\multirow{3}{*}{WordNet} & \multirow{3}{*}{146K} & \multirow{3}{*}{657K} & \multirow{3}{*}{4.50} & \texttt{kPathO} & 303 & 66 & 19 & 5 & 4 & 2 & 2 & 2 & 2 & 2\\
& & &  & \texttt{kPath1} & 294 & 67 & 18 & 4 & 3 & 3 & 1 & 1 & 1 & 1\\
& & &  & \texttt{kPath2} & 229 & 26 & 4 & 3 & 3 & 1 & 1 & 1 & 1 & 1\\
& & &  & \texttt{kPathN} & 4 & 3 & 3 & 2 & 2 & 1 & 1 & 1 & 1 & 1\\
& & &  & Red$\%$ & 98 & 95 & 84 & 60 & 50 & 50 & 50 & 50 & 50 & 50\\
\hline
\multirow{3}{*}{Douban} & \multirow{3}{*}{155K} & \multirow{3}{*}{327K} & \multirow{3}{*}{2.11} & \texttt{kPathO} & 284 & 7 & 3 & 2 & 2 & 2 & 2 & 2 & 2 & 2\\
& & &  & \texttt{kPath1} & 278 & 7 & 2 & 2 & 1 & 1 & 1 & 1 & 1 & 1\\
& & &  & \texttt{kPath2} & 214 & 2 & 2 & 1 & 1 & 1 & 1 & 1 & 1 & 1\\
& & &  & \texttt{kPathN} & 3 & 2 & 2 & 1 & 1 & 1 & 1 & 1 & 1 & 1\\
& & &  & Red$\%$ & 98 & 71 & 33 & 50 & 50 & 50 & 50 & 50 & 50 & 50\\
\hline
\multirow{3}{*}{Gowalla} & \multirow{3}{*}{197K} & \multirow{3}{*}{950K} & \multirow{3}{*}{4.83} & \texttt{kPathO} & 617 & 32 & 6 & 4 & 6 & 2 & 2 & 2 & 2 & 2\\
& & &  & \texttt{kPath1} & 603 & 28 & 6 & 5 & 2 & 2 & 1 & 1 & 1 & 1\\
& & &  & \texttt{kPath2} & 475 & 7 & 5 & 2 & 2 & 1 & 1 & 1 & 1 & 1\\
& & &  & \texttt{kPathN} & 4 & 2 & 2 & 2 & 2 & 1 & 1 & 1 & 1 & 1\\
& & &  & Red$\%$ & 99 & 93 & 66 & 50 & 66 & 50 & 50 & 50 & 50 & 50\\
\hline
\multirow{3}{*}{Dblp} & \multirow{3}{*}{317K} & \multirow{3}{*}{1.05M} & \multirow{3}{*}{3.31} & \texttt{kPathO} & 471 & 43 & 13 & 9 & 2 & 2 & 2 & 2 & 2 & 2\\
& & &  & \texttt{kPath1} & 453 & 47 & 14 & 6 & 5 & 1 & 1 & 1 & 1 & 1\\
& & &  & \texttt{kPath2} & 369 & 11 & 6 & 5 & 1 & 1 & 1 & 1 & 1 & 1\\
& & &  & \texttt{kPathN} & 4 & 3 & 3 & 3 & 1 & 1 & 1 & 1 & 1 & 1\\
& & &  & Red$\%$ & 99 & 93 & 76 & 66 & 50 & 50 & 50 & 50 & 50 & 50\\
\hline
\multirow{3}{*}{Amazon} & \multirow{3}{*}{355K} & \multirow{3}{*}{926K} & \multirow{3}{*}{2.76} & \texttt{kPathO} & 309 & 107 & 8 & 2 & 2 & 2 & 2 & 2 & 2 & 2\\
& & &  & \texttt{kPath1} & 322 & 100 & 7 & 4 & 1 & 1 & 1 & 1 & 1 & 1\\
& & &  & \texttt{kPath2} & 246 & 7 & 4 & 1 & 1 & 1 & 1 & 1 & 1 & 1\\
& & &  & \texttt{kPathN} & 4 & 3 & 2 & 1 & 1 & 1 & 1 & 1 & 1 & 1\\
& & &  & Red$\%$ & 98 & 97 & 75 & 50 & 50 & 50 & 50 & 50 & 50 & 50\\
\hline
Average  & -  & - & - & - & 90 & 82 & 63 & 56 & 55 & 51 & 50 & 50 & 48 & 46\\
\hline
\end{tabular}}
\caption{Improvement percentage for \texttt{kPathO} and \texttt{kPathN}  Algorithms}
\label{tab:KPathold_KPathnew_RedPer}
\end{table}

\begin{table}
\centering
\scalebox{0.75}{
    \begin{tabular}{ |p{1.3cm}|p{0.9cm}|p{0.9cm}|p{0.7cm} |p{1.3cm}|p{.8cm}| p{.5cm}|p{.5cm}|p{.5cm}|p{.5cm}|p{.5cm}| p{.5cm}|p{.5cm}|p{.5cm}|p{.5cm}|   }
    \hline
Dataset & $n$ & $m$ & $m/n$ & Heuristic & $k=1$ & $2$ & $3$ & $4$ & $5$ & $6$ & $7$ & $8$ & $9$ & $10$\\
\hline
\multirow{3}{*}{CU} & \multirow{3}{*}{49} & \multirow{3}{*}{107} & \multirow{3}{*}{2.18} & \texttt{kLevO} & 3 & 3 & 3 & 3 & 3 & 3 & 3 & 3 & 3 & 3\\
& & &  & \texttt{kLev1} & 3 & 3 & 3 & 3 & 3 & 3 & 3 & 3 & 3 & 2\\
& & &  & \texttt{kLev2} & 2 & 2 & 1 & 1 & 1 & 1 & 1 & 1 & 1 & 1\\
& & &  & \texttt{kLevN} & 1 & 1 & 1 & 1 & 1 & 1 & 1 & 1 & 1 & 1\\
& & &  & Red$\%$ & 66 & 66 & 66 & 66 & 66 & 66 & 66 & 66 & 66 & 66\\
\hline
\multirow{3}{*}{AJazz} & \multirow{3}{*}{198} & \multirow{3}{*}{2.74K} & \multirow{3}{*}{13.85} & \texttt{kLevO} & 3 & 3 & 3 & 3 & 3 & 3 & 3 & 3 & 3 & 3\\
& & &  & \texttt{kLev1} & 2 & 2 & 2 & 2 & 2 & 2 & 2 & 2 & 2 & 2\\
& & &  & \texttt{kLev2} & 2 & 2 & 2 & 2 & 2 & 2 & 2 & 2 & 2 & 2\\
& & &  & \texttt{kLevN} & 2 & 2 & 2 & 2 & 2 & 2 & 2 & 2 & 2 & 2\\
& & &  & Red$\%$ & 33 & 33 & 33 & 33 & 33 & 33 & 33 & 33 & 33 & 33\\
\hline
\multirow{3}{*}{HM} & \multirow{3}{*}{2.43K} & \multirow{3}{*}{16.6K} & \multirow{3}{*}{6.86} & \texttt{kLevO} & 4 & 4 & 4 & 4 & 4 & 4 & 4 & 4 & 4 & 4\\
& & &  & \texttt{kLev1} & 4 & 4 & 4 & 4 & 4 & 4 & 4 & 4 & 4 & 4\\
& & &  & \texttt{kLev2} & 3 & 3 & 3 & 2 & 2 & 2 & 2 & 1 & 1 & 1\\
& & &  & \texttt{kLevN} & 3 & 3 & 2 & 2 & 2 & 2 & 1 & 1 & 1 & 1\\
& & &  & Red$\%$ & 25 & 25 & 50 & 50 & 50 & 50 & 75 & 75 & 75 & 75\\
\hline
\multirow{3}{*}{ArxAP} & \multirow{3}{*}{18.8K} & \multirow{3}{*}{198K} & \multirow{3}{*}{10.55} & \texttt{kLevO} & 5 & 5 & 5 & 5 & 5 & 5 & 5 & 5 & 5 & 5\\
& & &  & \texttt{kLev1} & 4 & 4 & 4 & 4 & 4 & 4 & 4 & 4 & 4 & 4\\
& & &  & \texttt{kLev2} & 4 & 4 & 3 & 3 & 2 & 2 & 2 & 2 & 2 & 2\\
& & &  & \texttt{kLevN} & 4 & 4 & 3 & 3 & 2 & 2 & 2 & 2 & 2 & 2\\
& & &  & Red$\%$ & 20 & 20 & 40 & 40 & 60 & 60 & 60 & 60 & 60 & 60\\
\hline
\multirow{3}{*}{AsCaida} & \multirow{3}{*}{26.5K} & \multirow{3}{*}{53.4K} & \multirow{3}{*}{2.02} & \texttt{kLevO} & 4 & 4 & 4 & 4 & 4 & 4 & 4 & 4 & 4 & 4\\
& & &  & \texttt{kLev1} & 4 & 4 & 4 & 4 & 4 & 4 & 4 & 4 & 4 & 4\\
& & &  & \texttt{kLev2} & 2 & 2 & 1 & 1 & 1 & 1 & 1 & 1 & 1 & 1\\
& & &  & \texttt{kLevN} & 2 & 1 & 1 & 1 & 1 & 1 & 1 & 1 & 1 & 1\\
& & &  & Red$\%$ & 50 & 75 & 75 & 75 & 75 & 75 & 75 & 75 & 75 & 75\\
\hline
\multirow{3}{*}{BrightK} & \multirow{3}{*}{58.2K} & \multirow{3}{*}{214K} & \multirow{3}{*}{3.68} & \texttt{kLevO} & 5 & 5 & 5 & 5 & 5 & 5 & 5 & 5 & 5 & 5\\
& & &  & \texttt{kLev1} & 4 & 4 & 4 & 4 & 4 & 4 & 4 & 4 & 4 & 4\\
& & &  & \texttt{kLev2} & 3 & 2 & 2 & 1 & 1 & 1 & 1 & 1 & 1 & 1\\
& & &  & \texttt{kLevN} & 3 & 2 & 1 & 1 & 1 & 1 & 1 & 1 & 1 & 1\\
& & &  & Red$\%$ & 40 & 60 & 80 & 80 & 80 & 80 & 80 & 80 & 80 & 80\\
\hline
\multirow{3}{*}{LMocha} & \multirow{3}{*}{104K} & \multirow{3}{*}{2.19M} & \multirow{3}{*}{21.07} & \texttt{kLevO} & 4 & 4 & 4 & 4 & 4 & 4 & 4 & 4 & 4 & 4\\
& & &  & \texttt{kLev1} & 3 & 3 & 3 & 3 & 3 & 3 & 3 & 3 & 3 & 3\\
& & &  & \texttt{kLev2} & 3 & 3 & 3 & 3 & 3 & 3 & 3 & 2 & 2 & 2\\
& & &  & \texttt{kLevN} & 3 & 3 & 3 & 3 & 3 & 3 & 2 & 2 & 2 & 2\\
& & &  & Red$\%$ & 25 & 25 & 25 & 25 & 25 & 25 & 50 & 50 & 50 & 50\\
\hline
\multirow{3}{*}{FlickrE} & \multirow{3}{*}{106K} & \multirow{3}{*}{2.32M} & \multirow{3}{*}{21.87} & \texttt{kLevO} & 5 & 5 & 5 & 5 & 5 & 5 & 5 & 5 & 5 & 5\\
& & &  & \texttt{kLev1} & 4 & 4 & 4 & 4 & 4 & 4 & 4 & 4 & 4 & 4\\
& & &  & \texttt{kLev2} & 4 & 3 & 4 & 3 & 3 & 3 & 3 & 3 & 3 & 3\\
& & &  & \texttt{kLevN} & 4 & 4 & 4 & 3 & 3 & 3 & 3 & 3 & 3 & 3\\
& & &  & Red$\%$ & 20 & 20 & 20 & 40 & 40 & 40 & 40 & 40 & 40 & 40\\
\hline
\multirow{3}{*}{WordNet} & \multirow{3}{*}{146K} & \multirow{3}{*}{657K} & \multirow{3}{*}{4.50} & \texttt{kLevO} & 6 & 6 & 6 & 6 & 6 & 6 & 6 & 6 & 6 & 6\\
& & &  & \texttt{kLev1} & 5 & 5 & 5 & 5 & 5 & 5 & 5 & 5 & 6 & 5\\
& & &  & \texttt{kLev2} & 4 & 3 & 2 & 2 & 2 & 2 & 1 & 1 & 1 & 1\\
& & &  & \texttt{kLevN} & 4 & 3 & 2 & 2 & 2 & 1 & 1 & 1 & 1 & 1\\
& & &  & Red$\%$ & 33 & 50 & 66 & 66 & 66 & 83 & 83 & 83 & 83 & 83\\
\hline
\multirow{3}{*}{Douban} & \multirow{3}{*}{155K} & \multirow{3}{*}{327K} & \multirow{3}{*}{2.11} & \texttt{kLevO} & 4 & 4 & 4 & 4 & 4 & 4 & 4 & 4 & 4 & 4\\
& & &  & \texttt{kLev1} & 3 & 3 & 3 & 3 & 3 & 3 & 3 & 3 & 3 & 3\\
& & &  & \texttt{kLev2} & 2 & 2 & 1 & 1 & 1 & 1 & 1 & 1 & 1 & 1\\
& & &  & \texttt{kLevN} & 2 & 1 & 1 & 1 & 1 & 1 & 1 & 1 & 1 & 1\\
& & &  & Red$\%$ & 50 & 75 & 75 & 75 & 75 & 75 & 75 & 75 & 75 & 75\\
\hline
\multirow{3}{*}{Gowalla} & \multirow{3}{*}{197K} & \multirow{3}{*}{950K} & \multirow{3}{*}{4.83} & \texttt{kLevO} & 6 & 6 & 6 & 6 & 6 & 6 & 6 & 6 & 6 & 6\\
& & &  & \texttt{kLev1} & 5 & 5 & 5 & 5 & 5 & 5 & 5 & 5 & 5 & 5\\
& & &  & \texttt{kLev2} & 4 & 3 & 2 & 2 & 2 & 1 & 1 & 1 & 1 & 1\\
& & &  & \texttt{kLevN} & 4 & 2 & 2 & 2 & 1 & 1 & 1 & 1 & 1 & 1\\
& & &  & Red$\%$ & 33 & 66 & 66 & 66 & 83 & 83 & 83 & 83 & 83 & 83\\
\hline
\multirow{3}{*}{Dblp} & \multirow{3}{*}{317K} & \multirow{3}{*}{1.05M} & \multirow{3}{*}{3.31} & \texttt{kLevO} & 6 & 6 & 6 & 6 & 6 & 6 & 6 & 6 & 6 & 6\\
& & &  & \texttt{kLev1} & 4 & 4 & 4 & 4 & 4 & 4 & 4 & 5 & 5 & 5\\
& & &  & \texttt{kLev2} & 3 & 2 & 2 & 2 & 1 & 1 & 1 & 1 & 1 & 1\\
& & &  & \texttt{kLevN} & 3 & 2 & 2 & 1 & 1 & 1 & 1 & 1 & 1 & 1\\
& & &  & Red$\%$ & 50 & 66 & 66 & 83 & 83 & 83 & 83 & 83 & 83 & 83\\
\hline
\multirow{3}{*}{Amazon} & \multirow{3}{*}{355K} & \multirow{3}{*}{926K} & \multirow{3}{*}{2.76} & \texttt{kLevO} & 6 & 6 & 6 & 6 & 6 & 6 & 6 & 6 & 6 & 6\\
& & &  & \texttt{kLev1} & 5 & 5 & 5 & 5 & 5 & 5 & 5 & 5 & 5 & 5\\
& & &  & \texttt{kLev2} & 3 & 2 & 2 & 2 & 1 & 1 & 1 & 1 & 1 & 1\\
& & &  & \texttt{kLevN} & 3 & 2 & 2 & 1 & 1 & 1 & 1 & 1 & 1 & 1\\
& & &  & Red$\%$ & 50 & 66 & 66 & 83 & 83 & 83 & 83 & 83 & 83 & 83\\
\hline
Average  & -  & - & - & - & 38 & 49 & 56 & 60 & 63 & 64 & 68 & 68 & 68 & 68\\
\hline
\end{tabular}}
\caption{Improvement percentage for \texttt{kLevO} and \texttt{kLevN} Algorithms}
\label{tab:KLevold_KLevnew_RedPer}
\end{table}

% \begin{figure}[h!]
%   \begin{subfigure}{.5\textwidth}
%   \centering
%     \includegraphics[width=\linewidth]{arxiv_images/errorbar/errorbarklev_individual.png}
%   \end{subfigure}%
%   \begin{subfigure}{.5\textwidth}
%   \centering
%     \includegraphics[width=\linewidth]{arxiv_images/errorbar/errorbarkpath_individual.png}
%   \end{subfigure}
%   \caption{Improvement in required passes for \texttt{kLev} (left) and \texttt{kPath} (right) for real datasets .}
%   \label{fig:realindividualImprov}
% \end{figure}

%\clearpage
%\pagebreak

\section{Pseudocodes for final Algorithms}
\label{apn:pseudocodes}

For \texttt{kPath}, regarding the \texttt{H1} heuristic, previously, we used to call the Compute-DFS-\texttt{kPathO} from the second pass as the first pass is dedicated to the formation of the initial spamming tree and components by adding an artificial root. Now, as per our heuristic \texttt{H1}, the initial spamming tree is formed by adding an artificial root before 1st pass by adding edges to all vertices from the artificial root. Now, we can call the Compute-DFS-\texttt{kPathO} from the first pass itself. In computer-DFS-\texttt{kPathO}, $k_{opt}$ is the optimal value of k for storing saved edges of each component, which is calculated in every pass. In compute-DFS-\texttt{kLevN}, $k’$ represents the safe level, meaning that vertices with a level below it can be safely added at the end of the pass. $H_C$, in this case, stores the edges of top k’ levels along with Tc. $broomstick \textunderscore end \textunderscore vertex$ mentioned in the pseudocode represents the lowest vertex for which we need not store back edges, as explained in the heuristic \texttt{H3}. This value is initially set to artificial root and with passes, it drops to lower vertices.

Different colors in the pseudocodes represent different  implementations of heuristics: 
\begin{itemize}
    \item Red: Marked/Unmarked heuristic(\texttt{H0})
    \item Green:  Optimizing $nk$ edge utilization level 1 (\texttt{H2})
    \item Blue:  Optimizing $nk$ edge utilization level 2 (\texttt{H3})
\end{itemize}

\begin{procedure}
%\BlankLine
$E'_C\leftarrow \emptyset$\;
\tcc{Initiate a pass over $E$, for all components in parallel}
\textcolor{blue}{$k_{opt}$ = (n$\cdot$k)/no. of unvisited vertices\;}
\While(\tcc*[f]{process first $|V_C|(k-1)$ edges of $C$})
{\textcolor{blue}{$|E'_C|\leq |V_C|(k_{opt}-1)$} or Stream is over}
{
%\tcc{Let $e$ be the next edge in input stream}
\lIf{next edge $e$ belongs to $C$ and \textcolor{LimeGreen}{not belongs to $T_C$}}
{$E'_C\leftarrow E'_C\cup \{e\}$}
}

\BlankLine
\lIf{$par(r_C)\neq \phi$}{Add $(r_C,par(r_C))$ to $T$}
$T'_C\leftarrow$ DFS tree of $T_C\cup E'_C$ from root $r_C$\;
%\BlankLine
\lIf(\tcc*[f]{pass over $E$ was completed}){$|E'_C|\cup T_C\leq |V_C|\cdot k$}
{Add $T'_C$ to $T$}

\Else{
$P\leftarrow$ Path from $r_C$ to lowest vertex in $T'_C$\;
Add $P$ to $T$\;
\BlankLine
\tcc{Continue the pass for all components in parallel}
\ForAll{edges in $E'_C$ followed by the remaining pass over $E$}
{
%\ForAll{edges in a single pass over $E$}{
Compute the components $C_1,...,C_f$ of $C\setminus P$ using Union-Find algorithm\;
\tcc{This essentially computes $T_{C_1},...,T_{C_f}$}
Find lowest edge $e_i$ from each component $C_i$ to $P$\;
}

\ForEach{Component $C_i$ of $C\setminus P$}{
%Add $e_i$ to $T$\;
$par(y_i)\leftarrow$ $x_i$
\tcc*[r]{Let $e_i=(x_i,y_i)$, where $y_i\in C_i$}
\ref{alg:KPathN}($C_i$,$T_{C_i}$,$y_i$)\;
}

}
\caption{Compute-DFS-kPathN($C$,$T_C$,$r_C$): Computes a DFS tree of the 
%subgraph of $G$ induced by the 
component $C$ rooted at the vertex $r_C\in C$ along with the proposed heuristics.}
\label{alg:KPathN}
\end{procedure}

% \begin{procedure}
% %\BlankLine
% Initialize $H_C\leftarrow T_C$\; 

% $k' \leftarrow \infty$ \;

% Assign marked = 0 for all vertices of C\;

% $E_C\leftarrow \emptyset$ \;

% \tcc{Initiate a pass over $E$, for all components in parallel}
% \ForEach(\tcc*[f]{edges in the input stream })
% {edge $(x,y)$ in $E$ if $(x,y)\in E_C$}
% {
% {
% \If(){size of $E_C$ == $V_C$\cdot k}
% {

% remove an edge from $E_C$ connected to lowest vertex $u$\;
% $k'\leftarrow level(u)-1$\;

% }
% %\lIf(\tcc*[f]{ignore edge of lower component}){$rep[x]= rep[y]$}{Continue}
% Add $(x,y)$ to $H_C$\;
% %		\BlankLine
% %		\If{$(x,y)$ is a cross edge}
% %			{
% 			\ref{alg:rebuild}($T_C,(x,y)$)\;
% %			\ref{alg:rebuild}($x,y,T_C$)\;
% 			Remove excess edges from $H_C$
% 			\tcc*[r]{non-tree edges in $T_C\setminus T'_C$}
% %			Update $H_C$ and $rep[v],\forall v\in V_C$\;
% %			}	
% }
% }

% $T'_{C}\leftarrow$ Top $k$ levels of $T_C$
% \tcc*[r]{vertices $v$ with $0\leq level\leq k-1$}
% 			\BlankLine
% \lIf{$par(r_C)\neq \phi$}{Add $(r_C,par(r_C))$ to $T$}
% Add $T'_{C}$ to $T$\;

% 			\BlankLine

% \ForEach{tree $\tau \in T_C\setminus T'_C$}
% {
% $v \leftarrow root(\tau)$\;
% \tcc{Let $C_v$ be component containing $v$ in $C \setminus T'_C$}
% $par(v)\leftarrow$ Parent of $v$ in $T_C$\;
% \ref{alg:KLevN}($C_v$,$T_C(v)$,$v$):
% }

% \caption{Compute-DFS-KLevN($C$,$T_C$,$r_C$): Computes a DFS tree of the 
% component $C$ whose spanning tree $T_C$ is rooted at the vertex $r_C\in C$.}
% \label{alg:KLevN}
% \end{procedure}

\begin{procedure}
%\BlankLine
Initialize $H_C\leftarrow T_C$\; %, Compute $rep[v],\forall v\in V_C$\;

\textcolor{red}{$k' \leftarrow \infty$}\;

\textcolor{LimeGreen}{Assign marked $=$ 0 for all vertices of C}\;

\textcolor{red}{$E_C'\leftarrow \emptyset$} \;

\tcc{Initiate a pass over $E$, for all components in parallel}
\ForEach(\tcc*[f]{edges in the input stream from $C$})
{edge $(x,y)$ in $E$ if $(x,y)\in E_C$}
{

\If() {size of $E_C' == V_C \cdot k$}
{
\textcolor{red}{remove an edge from $E_C'$ connected to a vertex $u$, which is the farthest from the root}\;
\textcolor{red}{$k' \leftarrow level(u)-1$}\;
}

%\lIf(\tcc*[f]{ignore edge of lower component}){$rep[x]= rep[y]$}{Continue}
		
		Add $(x,y)$ to $H_C$\;
%		\BlankLine
%		\If{$(x,y)$ is a cross edge}
%			{
			\ref{alg:rebuild}($T_C,(x,y)$)\;

   \textcolor{blue}{Check whether $broomstick\textunderscore end\textunderscore vertex$ fall further or not}\;

   \textcolor{blue}{Remove back edges stored for vertices above $broomstick \textunderscore end \textunderscore vertex$}\;
%			\ref{alg:rebuild}($x,y,T_C$)\;
			Remove excess edges from $H_C$ and $E_C'$ \;

    \textcolor{LimeGreen}{Mark all vertices whose level has changed and are outside the top $k'$ levels}\;
%			Update $H_C$ and $rep[v],\forall v\in V_C$\;
%			}	
		
}

$T'_{C}\leftarrow$ \textcolor{LimeGreen}{All vertices with marked status $=$ 0}\;
			\BlankLine
\lIf{$par(r_C)\neq \phi$}{Add $(r_C,par(r_C))$ to $T$}
Add $T'_{C}$ to $T$\;

			\BlankLine

\ForEach{tree $\tau \in T_C\setminus T'_C$}
{
$v \leftarrow root(\tau)$\;
\tcc{Let $C_v$ be component containing $v$ in $C \setminus T'_C$}
$par(v)\leftarrow$ Parent of $v$ in $T_C$\;
\ref{alg:KLevN}($C_v$,$T_C(v)$,$v$):
}

\caption{Compute-DFS-kLevN($C$,$T_C$,$r_C$): Computes a DFS tree of the 
component $C$ whose spanning tree $T_C$ is rooted at the vertex $r_C\in C$.}
\label{alg:KLevN}
\end{procedure}	

\end{document}